\documentclass[a4paper,fleqn,usenatbib]{mnras}
\pdfoutput=1
\usepackage{newtxtext,newtxmath}

\usepackage[T1]{fontenc}
\usepackage{ae,aecompl}
\usepackage{threeparttable}
\usepackage{hyperref}

\usepackage{graphicx}	
\usepackage{amsmath}	
\usepackage{amssymb}	

\newcommand{\msol}{M$_{\odot}$}

\newcommand{\cgs}{erg s$^{-1}$ cm$^{-2}$}

\newcommand{\ot}{1$\times$3 }
\newcommand{\muv}{$M_{\mathrm{UV}}$ }

\title[Serendipitous Emission Line Galaxies with NIRSpec]{Abundant Serendipitous Emission Line Sources with JWST/NIRSpec}

\author[Maseda et al.] {Michael V. Maseda$^{1}$\thanks{E-mail: maseda@strw.leidenuniv.nl}, Marijn Franx$^{1}$, Jacopo Chevallard$^{2}$, and Emma Curtis-Lake$^{3}$ \\
  $^{1}$Leiden Observatory, Leiden University, P.O. Box 9513, 2300 RA, Leiden, The Netherlands \\ 
  $^{2}$Sorbonne Universit{\'e}s, UPMC-CNRS, UMR7095, Institut d'Astrophysique de Paris, F-75014 Paris, France\\
  $^3$Kavli Institute for Cosmology, Madingley Road, Cambridge CB3 0HA, UK}
  
\pubyear{2019}

\begin{document}
\label{firstpage}
\pagerange{\pageref{firstpage}--\pageref{lastpage}}
\maketitle

\begin{abstract}
The James Webb Space Telescope will provide observational capabilities that far exceed those of current ground- or space-based instrumentation.  In particular, the NIRSpec instrument will take highly sensitive spectroscopic data for hundreds of objects simultaneously from 0.6$-$5.3 $\mu$m.  Current photometric observations suggest a large and increasing number of faint (\muv $> -$16) galaxies at high-redshift, with increasing evidence that galaxies at these redshifts have optical emission lines with extremely high equivalent widths.  A simple model of their emission line fluxes and number density evolution with redshift is used to predict the number of galaxies that NIRSpec will serendipitously observe during normal observations with the microshutter array.  At exposure times of $\approx$ 20 hours in the low-resolution prism mode, the model predicts that, on average, every open \ot `microslit' will contain an un-targeted galaxy with a detectable \hbox{[O\,{\sc iii}]} and/or {\hbox{H $\alpha$}} emission line; while most of these detections are predicted to be of \hbox{[O\,{\sc iii}]}, {\hbox{H $\alpha$}} detections alone would still number 0.56 per open `microslit' for this exposure time.  Many of these objects are spectroscopically detectable even when they are fainter than current photometric limits and/or their flux centroids lie outside of the open microshutter area.  The predicted number counts for such galaxies match $z\sim2$ observations of \hbox{[O\,{\sc iii}]} emitters from slitless grism spectroscopic surveys, as well as theoretical predictions based on sophisticated modeling of galaxy spectral energy distributions.  These serendipitous detections could provide the largest numbers of $z>6$ spectroscopic confirmations in the deepest NIRSpec surveys.
\end{abstract}


\section{INTRODUCTION}

The James Webb Space Telescope \citep[JWST;][]{2006SSRv..123..485G} represents the most significant new space-based observatory at optical and near-infrared wavelengths of the current decade.  Currently scheduled to launch by March 2021\footnote{\url{https://www.nasa.gov/sites/default/files/atoms/files/webb_irb_report_and_response_0.pdf}}, JWST's spectroscopic capabilities from 0.6 to 28.8 $\mu$m far exceed those of existing observatories, such as the \textit{Hubble} Space Telescope (HST) or the \textit{Spitzer} Space Telescope.  In particular, the Near-Infrared Spectrograph \citep[NIRSpec;][]{2007SPIE.6692E..0MB,2010SPIE.7731E..0DB} will give us unprecedented spectral sensitivity and multi-object capabilities (in the multi-object spectroscopy or `MOS' mode) with uninterrupted coverage from 0.6$-$5.3 $\mu$m.  

This red wavelength coverage is required to detect bright restframe-optical emission lines, such as {\hbox{H $\alpha$}} and \hbox{[O\,{\sc iii}]} $\lambda$5007, at high-$z$ since current ground-based observations are typically limited to the $K$-band which extends out to $\sim$2.4 $\mu$m and hence $z\lesssim4$.  JWST/NIRSpec will detect these lines out to $z\sim9.6$ with one to two orders of magnitude higher spectroscopic sensitivity than existing instruments.  With the configurable microshutter array (MSA), NIRSpec can deliver up to $\sim$200 non-overlapping spectra simultaneously. This will be crucial in obtaining spectroscopic confirmations of the numerous high-$z$ galaxy candidates that have so far been detected primarily through HST imaging \cite[e.g.][]{2003MNRAS.342..439S,2004MNRAS.355..374B,2004ApJ...600L..99D,2017ApJ...843..129B,2017ApJ...835..113L,2018MNRAS.479.5184A,2018ApJ...854...73I}.

Recent observations have pointed to an abundant population of faint (\muv $>$ $-$18) $z\gtrsim4$ galaxies that are often difficult to study spectroscopically due to the intrinsic faintness of the UV (metal) emission lines and the absorption of the powerful {\hbox{Ly $\alpha$}} line from the neutral intergalactic medium for galaxies in the reionization epoch \citep{2014AA...569A..78V}.  However, photometric techniques have been developed to study the restframe-optical emission line properties of these galaxies: the existence of bright, high-equivalent width (EW) optical emission lines in $z\gtrsim4$ galaxies is inferred by measuring strong excesses in broad-band Spitzer/IRAC 3.6 $\mu$m and/or 4.5 $\mu$m photometry \cite[e.g.][]{2011ApJ...738...69S,2012ApJ...755..148G,2013ApJ...777L..19L,2013ApJ...763..129S,2014ApJ...784...58S}.  With EWs of \hbox{[O\,{\sc iii}]} and {\hbox{H $\alpha$}} often in excess of 500 \AA \ (restframe), these lines should be readily detectable with moderately-deep NIRSpec observations.


Obtaining large spectroscopic samples of $z\gtrsim4$ galaxies is one of the primary goals of NIRSpec.  These samples will be crucial in understanding galaxy formation and evolution, from the contribution of galaxies to the reionization of the universe \cite[e.g.][]{2012ApJ...752L...5B,2012ApJ...758...93F,2013ApJ...768...71R} and the relationship between galaxies and accreting supermassive black holes \cite[e.g.][]{2018MNRAS.473.2378V}, to the build-up of the stellar mass of the universe \cite[e.g.][]{2014MNRAS.444.2960D} and the chemical enrichment histories of the star-forming population \cite[e.g.][]{2008AA...488..463M}.  Even when we have access to exquisite JWST/NIRCam imaging, precise spectroscopic redshifts will be necessary to fully interpret a galaxy's spectral energy distribution \citep[SED; e.g.][]{2013MNRAS.435.2885W}.  If `serendipitous' sources that are not the primary target of observations contribute to the number counts of a NIRSpec survey, then NIRSpec becomes an even more powerful tool to assemble these large samples.

Using integral field unit (IFU) spectroscopy, \citet{2017AA...608A...3B} show that faint galaxies often have (spectroscopically-detectable) `contaminants' located nearby in projection: between 5 and 10 per cent of galaxies with $F775W$ magnitudes between 25 and 26 have a projected companion within 0\farcs5 that has a brighter emission line than the strongest line in the primary galaxy.  While this was discussed in the context of contamination in photometric versus spectroscopic redshift surveys, \citet{2017AA...608A...3B} find that spectroscopic observations with a spatial resolution of 0\farcs5 will be contaminated by an object with stronger emission lines 1 per cent of the time.  Even though high-$z$ sources are expected to be small in size \cite[$r_e \sim 0\farcs1$ at $z=6$;][]{2015ApJS..219...15S,2016MNRAS.457..440C}, the large and under-sampled NIRSpec point spread function (PSF), particularly at wavelengths $>$ 3 $\mu$m, also means that flux from these objects will be dispersed onto the detector even if they lie outside the 0\farcs20 $\times$ 0\farcs46 open area of each microshutter.  In addition, many of these sources could have spectroscopically-detectable emission lines even with continuum magnitudes below imaging detection limits \cite[cf. ][]{2001ApJ...560L.119E,2008ApJ...681..856R,2011AA...525A.143C,2012ApJ...744..149H,2017AA...608A...1B,2018ApJ...865L...1M}, making their characterisation with NIRSpec even more crucial.

Given the sensitivity of NIRSpec, the size of the open area of each microshutter, and the number of simultaneous spectra that can be obtained (up to 200 per configuration), we would therefore expect a significant number of `contaminating' spectra in a NIRSpec observation, even more so when coupled with the aforementioned evolution in the highest-EW optical emission lines with redshift.  This also extends beyond the estimation of the number of contaminants with stronger emission lines: the primary concern is the number of spectroscopically-detectable emission line sources, agnostic as to whether they are brighter than the primary spectroscopic target. 

To this end, we develop a model, based on continuum UV luminosity functions (Section \ref{sec:uvlf}), for the evolution of emission line fluxes ({\hbox{H $\alpha$}} and \hbox{[O\,{\sc iii}]}) with redshift (Section \ref{sec:lines}).  With a realistic model for NIRSpec multi-object observations (Section \ref{sec:nirspec}), we estimate the number of `serendipitous emission line sources' that are detectable as a function of observing mode and observing time (Section \ref{sec:numbers}).  Caveats and the interpretation of the results from the model are also presented (Section \ref{sec:discussion}).  `Microshutter' refers to a single 0\farcs20 $\times$ 0\farcs46 MSA shutter while `microslit' refers to a 1$\times$3 configuration of open microshutters (1\farcs52 tall, including the bars between shutters).  We adopt a flat $\Lambda$CDM cosmology ($\Omega_m=0.3$, $\Omega_\Lambda=0.7$, and H$_0=70 ~$km s$^{-1}$ Mpc$^{-1}$) and AB magnitudes \citep{1974ApJS...27...21O} throughout.

\section{UV Luminosity Functions}
\label{sec:uvlf}
Galaxy luminosity functions are traditionally fit with a Schechter function \citep{1976ApJ...203..297S} of the form:
\begin{equation}
\label{schechter}
n(M) \ \mathrm{d}M = 0.4 \ \ln 10 \ \phi^{\star}  [ 10^{ 0.4 ( M^{\star} - M ) } ]^{ \alpha + 1}  e^{[ -10^{ 0.4 ( M^{\star} - M ) } ]} \ \mathrm{d}M, 
\end{equation}
describing the number of galaxies $n$ of a given magnitude $M$, where $\phi^{\star}$ is the normalisation (units of Mpc$^{-3}$), $M^{\star}$ is the characteristic magnitude, and $\alpha$ is the faint-end power law slope.  An equivalent form of this function can be written in terms of luminosity $L$ instead of $M$.

\begin{figure}
\begin{center}
\includegraphics[width=.45\textwidth]{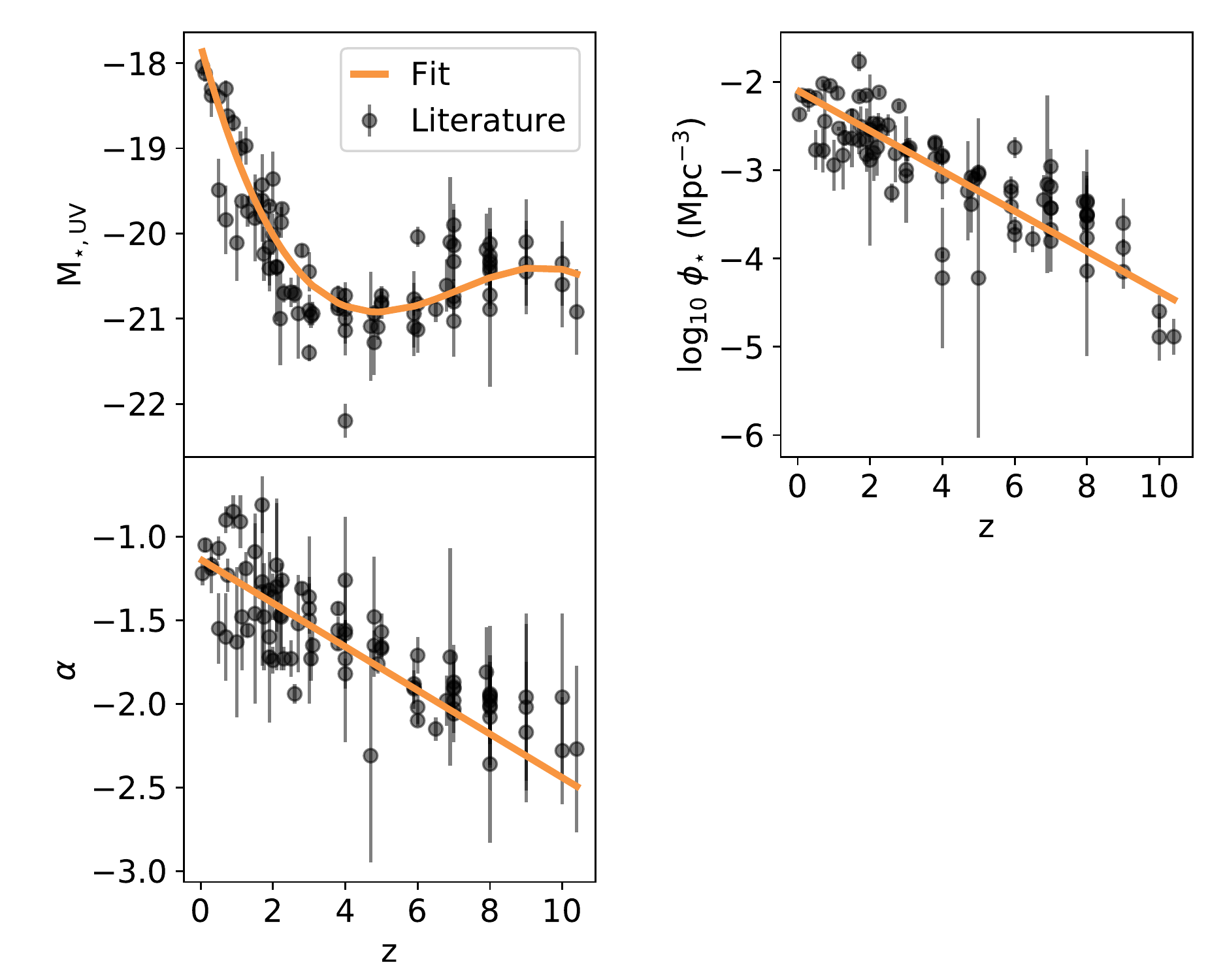}
\end{center}
\caption{Literature data (Table \ref{tab:uvlf}) and polynomial fits to $\alpha$, M$_{\star,\mathrm{UV}}$, and log$_{10}~\phi_{\star}$ as a function of redshift.  For M$_{\star,\mathrm{UV}}$, the fit is a third-order polynomial to capture the flattening of the relation at high-z \citep{2015ApJ...803...34B} and is measured at $\sim$1550 \AA.  The results of the fits are given in Equations \ref{eqn1}, \ref{eqn2}, and \ref{eqn3}.}
\label{fig:polyfit}
\end{figure}

Predictions for the volume density of galaxies with a given \muv have been made from the local universe out to the highest observed redshifts.  Here, we combine results from many studies to determine a single UV luminosity function parametrized by redshift \citep[see also][]{2016MNRAS.456.3194P,2018ApJS..236...33W}.  We fit linear functions to the Schechter function parameters $\alpha$(z) and log$_{10}$ $\phi_{\star}$(z), and a fourth-order polynomial to $M_{\star,\mathrm{UV}}$($z$) to reproduce the asymptotic behavior at high-$z$ \citep{2015ApJ...803...34B}; see Figure \ref{fig:polyfit}.  These parameters are listed in Table \ref{tab:uvlf}.  The $M_{\star,\mathrm{UV}}$ values from the literature span a range in restframe wavelength from 1500 to 1700 \AA$~$which we consider to be $M_{\star,1550}$ as the average galaxy in these samples has a negligible $K$-correction from these wavelengths to 1550 \AA$~$\citep{2009ApJ...690.1350O}.  The best-fit parameterizations are:
\begin{equation}
\label{eqn1}
\alpha(z) = -0.131 \times z -1.14,
\end{equation}
\begin{equation}
\label{eqn2}
log_{10}~\phi_{\star}(z) = -0.228 \times z -2.10,
\end{equation}
and
\begin{equation}
\label{eqn3}
M_{\star}(z) = -0.0112 \times z^3 + 0.241 \times z^2 -1.55 \times z - 17.8.
\end{equation}

The faint-end slope of the galaxy UV luminosity function has recently received considerable attention.  In particular, at $z\sim6$ many authors have used data from the Hubble Frontier Fields \citep{2017ApJ...837...97L}, leveraging gravitational lensing to find extremely faint galaxies.  As mentioned before, the interest in the number density of such faint galaxies is due to the potential contribution of such faint (\muv $> -$16) galaxies to cosmic reionization.  However, the numbers of photometric candidates at these faint limits are small and large uncertainties on the lensing magnification and the redshift of the sources remain.

\begin{figure}
\begin{center}
\includegraphics[width=.45\textwidth]{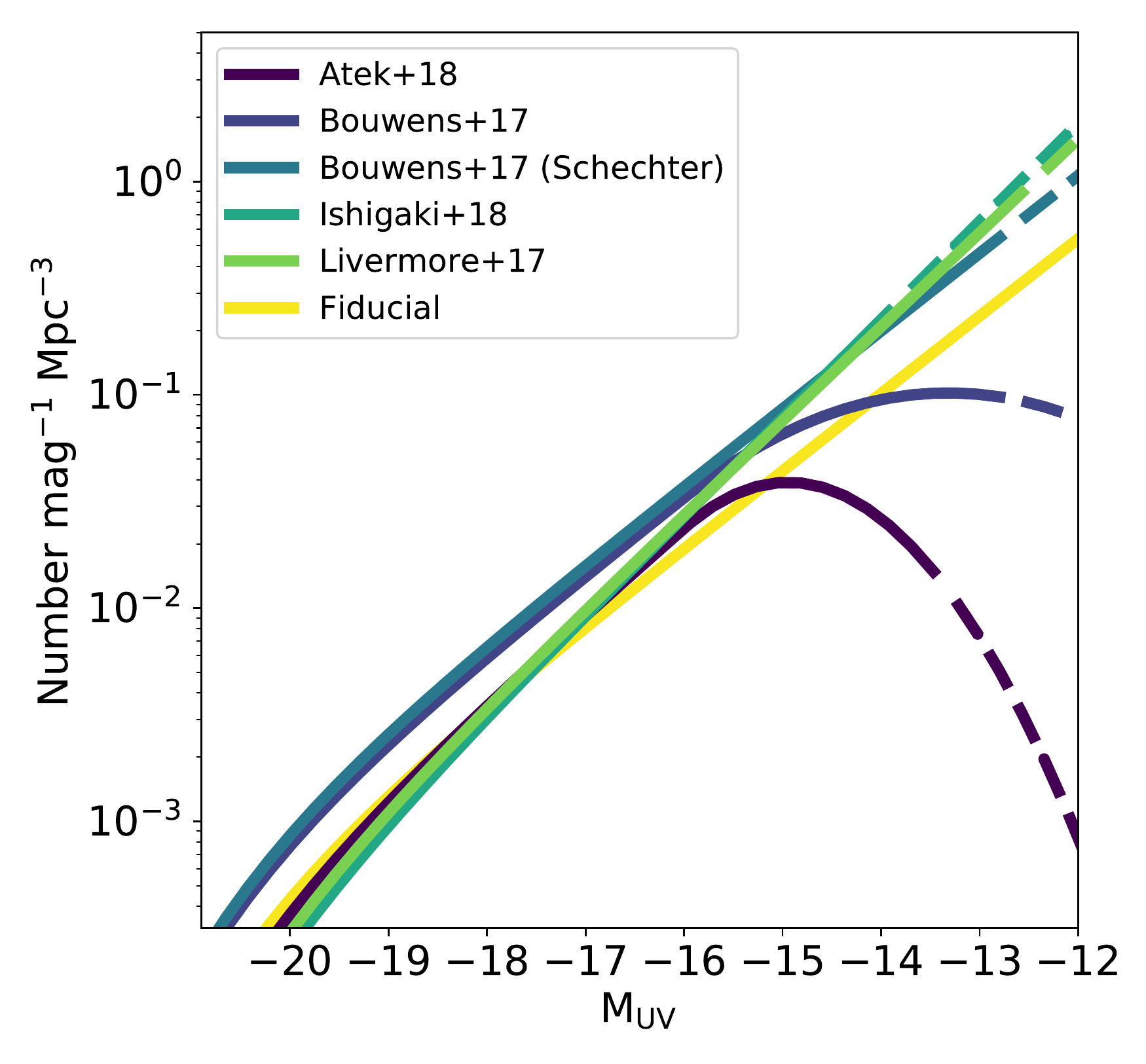}
\end{center}
\caption{The faint end of the $z\sim6$ UV luminosity function from \citet{2018MNRAS.479.5184A}, \citet{2017ApJ...843..129B}, \citet{2018ApJ...854...73I}, and \citet{2017ApJ...835..113L}, as well as our fiducial luminosity function (Section \ref{sec:uvlf}).  In all cases except for the fiducial model, the dashed region of the curves denotes an extrapolation to the observed trend at brighter magnitudes.  \citet{2018MNRAS.479.5184A} and \citet{2017ApJ...843..129B} fit for a declining luminosity function above \muv $= -$16 \citep[also shown is the Schechter function fit to][]{2017ApJ...843..129B}.  Given their similarity, we will consider the \citet{2018ApJ...854...73I} and \citet{2017ApJ...835..113L} results together throughout.  As discussed in Sections \ref{sec:uvlf} and \ref{sec:nonschechter}, the shape of the faint end of the luminosity function can have a strong effect on the results presented in this work.}

\label{fig:z6}
\end{figure}

    Based on the existing data, some authors have suggested that deviations from a power-law slope exist at \muv fainter than $-$16 \citep{2017ApJ...843..129B,2018MNRAS.479.5184A}, while others find good agreement with a power-law even at the faintest magnitudes probed \citep[e.g.][]{2017ApJ...835..113L,2018ApJ...854...73I}; see Figure \ref{fig:z6}.  This is in contrast to results at lower-redshifts, where a power-law slope is observed at least until \muv = $-$13 \citep{2016ApJ...832...56A}.

Throughout the rest of this work, we will consider the fiducial (Schechter) luminosity function derived above at all redshifts.  However, in Section \ref{sec:nonschechter}, we will explore the effect of these different $z\sim6$ luminosity functions on the number of detectable serendipitous emission line sources.

\section{Relationship Between \muv and Emission Lines}
\label{sec:lines}
\begin{figure}
\begin{center}
\includegraphics[width=.45\textwidth]{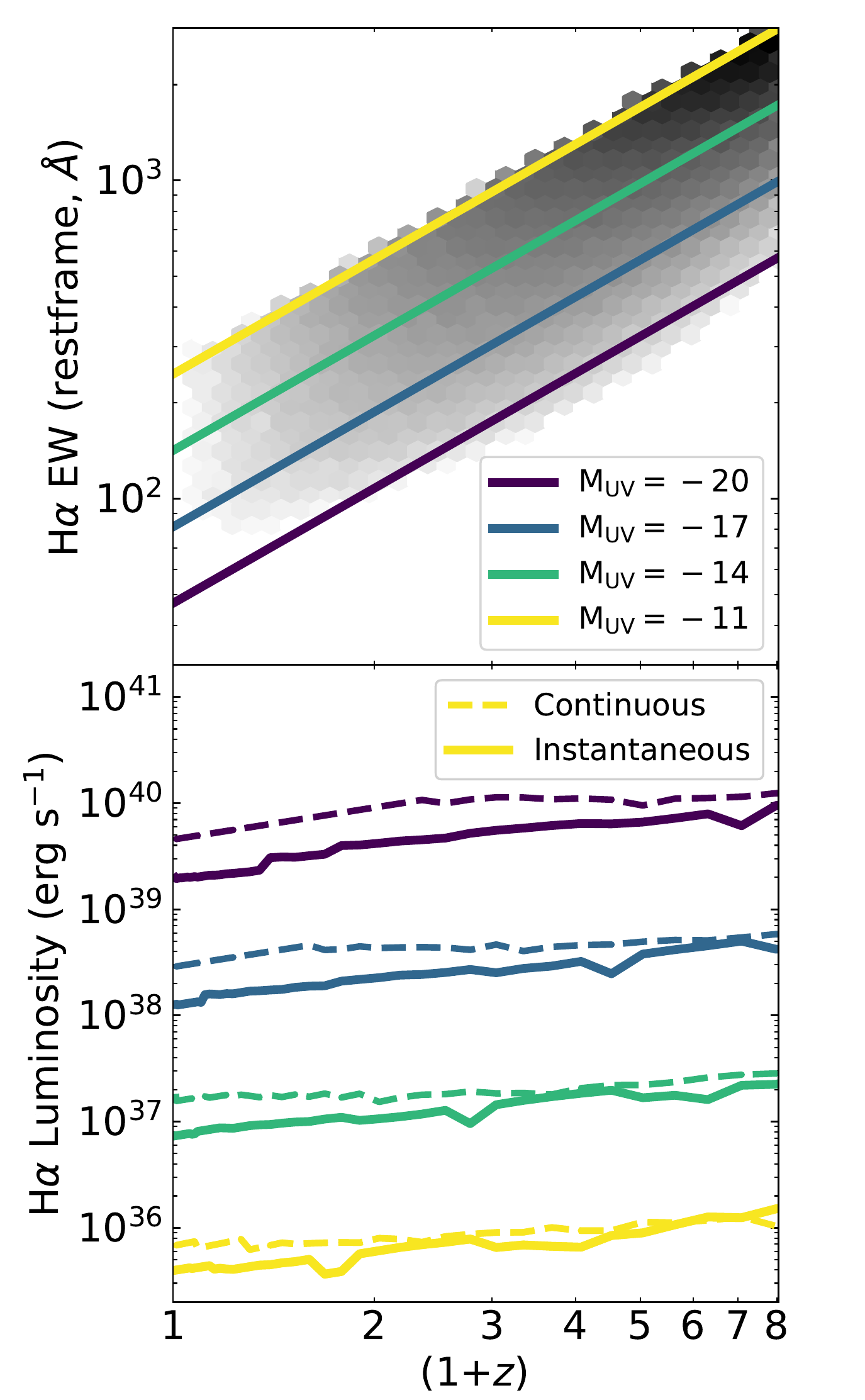}
\end{center}
\caption{(Top) Relationship between EW$_{H\alpha}$, $M_{\mathrm{UV}}$, and redshift from our model (Equation \ref{eqn:ew}).  The background shading shows the (log) density of values for galaxies predicted from our fiducial UV luminosity function (for clarity, only galaxies with \muv $< -$11 are shown).  (Bottom) Predicted {\hbox{H $\alpha$}} luminosity versus redshift for the same \muv values.  Our fiducial model with an instantaneous burst of star formation is shown with the solid line while a model using continuous star formation (see Section \ref{sec:sfh} for details) is shown with a dashed line.}
\label{fig:ew}
\end{figure}

While (UV continuum) luminosity functions predict the number density of galaxies at a given redshift, they do not provide direct constraints on the strength of emission lines, which we require to estimate the number of spectroscopic detection of galaxies.  While various studies have concluded that the `average' galaxy at $z \gtrsim 4-6$ has high-EW nebular emission lines \citep[e.g.][]{2012ApJ...755..148G,2013ApJ...763..129S,2013ApJ...777L..19L,2019MNRAS.484.1366C}, \citet{2016ApJ...833..254S} derive a relationship between $M_{\mathrm{UV}}$ and EW$_{0,H\alpha}$ at $z\sim4.4$.  At these redshifts, the strength of {\hbox{H $\alpha$}} (plus \hbox{[N\,{\sc ii}]} and \hbox{[S\,{\sc ii}]}) is determined via a photometric excess in the \textit{Spitzer}/IRAC photometry.  The relation from \citet{2016ApJ...833..254S} is shallow, with a slope $d$ log$_{10}~EW_{0,H\alpha}/d M_{UV}$ of 0.08.  To account for the redshift evolution of this relation, we use the result from \citet{2013ApJ...777L..19L}, who found that EW$_{0,H\alpha}$ increases as $(1+z)^{1.2}$ out to $z\sim8$.  Therefore, we parametrize EW$_{0,H\alpha}$ according to:
\begin{equation}
EW_{0,\mathrm{H\alpha}} (M_{\mathrm{UV}}, z) = 10^{(0.08\times M_{\mathrm{UV}} + 4.15)}\left(\frac{1+z}{1+4.4}\right)^{1.2}.
\label{eqn:ew}
\end{equation}
We also incorporate the observed scatter from \citet{2016ApJ...833..254S} into the relation, assuming a Gaussian scatter of $\sigma$ = 0.2 dex.  For certain redshifts and \muv values, the EW value predicted by Equation \ref{eqn:ew} exceeds 3000 \AA, the maximum predicted value from \texttt{Starburst99} at $Z=0.004$ ($\sim$ 0.25 Z$_{\odot}$).  In such cases, we adopt this value for EW.  This relation is shown in Figure \ref{fig:ew}.

While this relationship allows us to predict the EW of {\hbox{H $\alpha$}}, this does not directly translate into an observable {\hbox{H $\alpha$}} flux.  Obtaining a line flux from an EW requires knowledge of the continuum level at the position of the line, and the relation between $M_{UV,1550}$ and $M_{6563}$ depends on the stellar population properties.  We therefore adopt the models of \texttt{Starburst99} \citep{1999ApJS..123....3L} to convert EW$_{0,H\alpha}$ to $f_{H\alpha}$: we assume an instantaneous burst of star formation with a metallicity $Z=0.004$ and a \citet{1955ApJ...121..161S} initial mass function sampled between 1 and 100 M$_{\odot}$.  With these models, EW$_{0,H\alpha}$ maps nearly monotonically to a starburst age.  Using this {\hbox{H $\alpha$}}-derived value for the age of the starburst, the emergent spectrum from the model, normalised to the same $M_{UV,1550}$ as the observed galaxy, is used to estimate the local continuum level at the position of {\hbox{H $\alpha$}}.  This continuum level is then used to convert EW$_{0,H\alpha}$ into a line flux.

We would also like to predict the flux of other optical emission lines, namely \hbox{[O\,{\sc iii}]} $\lambda$5007.  The ratio of {\hbox{H $\alpha$}} to \hbox{[O\,{\sc iii}]} primarily depends on the metallicity and the ionization parameter and can be as large as 5 in the fiducial photoionization model of \citet{2016MNRAS.462.1757G} at $Z=0.004$.  In addition, near-IR grism spectroscopic observations of $1.1 < z < 2.4$ galaxies with restframe \hbox{[O\,{\sc iii}]} EWs in excess of 500 \AA \ by \citet{2018ApJ...854...29M}, the so-called `Extreme Emission Line Galaxies' \citep{2011ApJ...743..121A,2011ApJ...742..111V}, have an observed (\hbox{[N\,{\sc ii}]}-corrected) \hbox{[O\,{\sc iii}]}/{\hbox{H $\alpha$}} ratio of 2.1 \citep[cf. the value of 1.7 from][]{2003AA...401..1063A}.  These are precisely the types of galaxies that we expect to dominate the population at faint magnitudes and high redshifts and hence dominate the population of serendipitous emission line sources.  Therefore, we adopt a value of $f_{\hbox{[O\,{\sc iii}]}}$/$f_{H\alpha}=$ 2.1 $\pm$ 1 (normally-distributed).  Further discussion on this point is provided in Section \ref{sec:oiii}.  For {\hbox{H $\beta$}}, we assume case B recombination and hence $f_{H\alpha}$/$f_{H\beta}=2.86.$ While we perform the same analysis for {\hbox{H $\beta$}} as we do for {\hbox{H $\alpha$}} and include its effects in e.g. the contribution to broadband magnitudes in Section \ref{sec:imaging}, we do not consider {\hbox{H $\beta$}}-based detections given its wavelength proximity to the brighter \hbox{[O\,{\sc iii}]} emission line.

As we do not consider emission lines apart from \hbox{[O\,{\sc iii}]} and {\hbox{H $\alpha$}}, our results represent a lower limit to the number of potentially observable sources.  Another restframe-optical emission line, \hbox{[O\,{\sc ii}]} $\lambda\lambda$3727,3729, is commonly observed in star-forming galaxies and would be detectable in NIRSpec up to $z\sim13$ (currently beyond the constraints from UV luminosity functions based on HST imaging).  The ratio of \hbox{[O\,{\sc iii}]} to \hbox{[O\,{\sc ii}]} varies with ionization parameter, stellar mass, star formation rate, and metallicity \citep{2014MNRAS.442..900N}.  As such, $z<1$ observations of star-forming galaxies show a mean ratio of 0.4 \citep{2018arXiv180804899P} whereas the low-mass, low-metallicity $z\sim2$ `Extreme Emission Line Galaxies' (similar to the sources we expect to dominate the serendipitous NIRSpec counts) have a mean ratio of 3.5 \citep{2018ApJ...854...29M}, with evidence that the ratio increases with \hbox{[O\,{\sc iii}]} EW \citep{2018arXiv180909637T}.  Hence, large samples of \hbox{[O\,{\sc ii}]} emitters might be difficult to obtain at high-$z$.  The \citet{2018ApJS..236...33W} JAGUAR catalogue predicts 1/2 as many detectable \hbox{[O\,{\sc ii}]} emitters as \hbox{[O\,{\sc iii}]} emitters (see Section \ref{sec:jaguar}). {\hbox{Ly $\alpha$}} is also strong in many star-forming galaxies, potentially stronger than {\hbox{H $\alpha$}} \citep[e.g. in $z\sim$ 2, 10$^{10}$ M$_{\odot}$ galaxies;][]{2016MNRAS.458..449M}.  However, it is difficult to relate its strength to \muv in general \citep[e.g.][]{2006ApJ...645L...9A,2009MNRAS.400..232N} and it may not be easily observable at $z>6$ where the photons would be absorbed by the predominantly neutral intergalactic medium \cite[e.g.][]{2011ApJ...743..132P,2012ApJ...747...27T}.

\section{NIRSpec Observations}
NIRSpec offers nine unique spectral configurations spanning the wavelength range 0.6$-$5.3 $\mu$m. This wavelength interval is divided into four wavelength bands (henceforth Bands \hbox{{\sc 0}}, \hbox{{\sc i}}, \hbox{{\sc ii}}, and \hbox{{\sc iii}}) which are selected via different long-pass filters: $F070LP$, $F100LP$, $F170LP$, and $F290LP$.  In each band, specific diffraction gratings can provide $R\sim$ 1000 spectral resolution ($G140M$, $G235M$, and $G395M$) and $R\sim$ 2700 ($G140H$, $G235H$, and $G395H$); Bands \hbox{{\sc 0}} and \hbox{{\sc i}} can be used with either of the G140 gratings. Therefore, Band \hbox{{\sc 0}} covers 0.7 $\mu$m $< \lambda <$ 1.3 $\mu$m, Band \hbox{{\sc i}} covers 1.0 $\mu$m $< \lambda <$ 1.8 $\mu$m, Band \hbox{{\sc ii}} covers 1.7 $\mu$m $< \lambda <$ 3.1 $\mu$m, and Band \hbox{{\sc iii}} covers 2.9 $\mu$m $< \lambda <$ 5.3 $\mu$m.  NIRSpec's complete wavelength coverage can be achieved simultaneously with the $R\sim$ 100 prism (and the `CLEAR' blocking filter).  A more complete description of NIRSpec's observation modes can be found in \citet{nirspec}.

\label{sec:nirspec}
\subsection{Instrument sensitivity, slit losses, and spectral resolution}

To model the NIRSpec observations we use version \texttt{1.3} of the \texttt{Pandeia} \citep{10.1117/12.2231768} code, which is a scriptable exposure time calculator.   \texttt{Pandeia} is based on ground test measurements and calibrations of each of JWST's science instruments. for NIRSpec, \citet{10.1117/12.2231768} use data from extensive ground-based cryogenic testing of the instrument science module.  They do, however, caution that some uncertainties about the overall throughputs and efficiencies will remain until on-orbit calibrations are performed.  

\begin{figure*}
\begin{center}
\includegraphics[width=.9\textwidth]{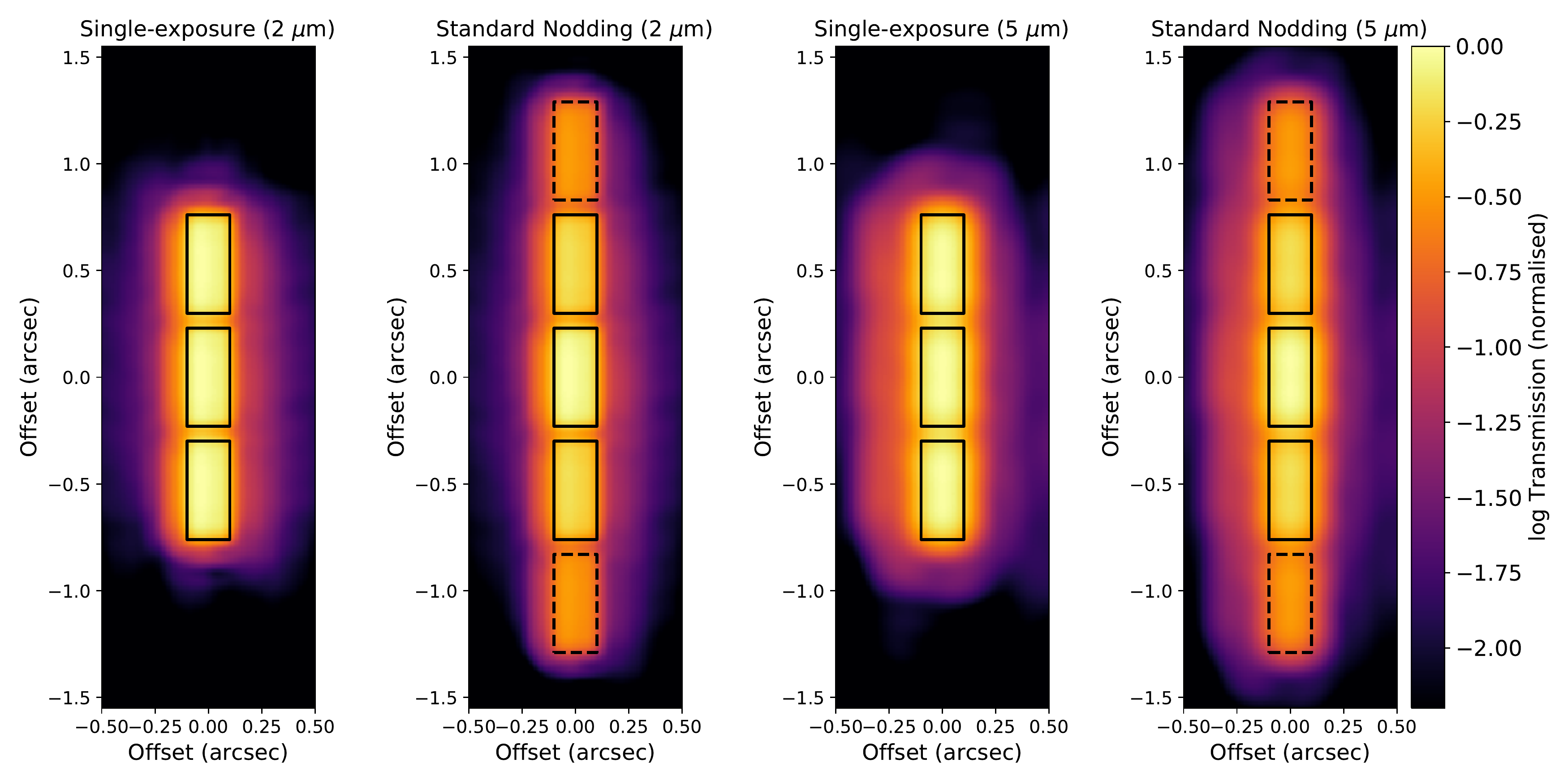}
\end{center}
\vspace{-0.5cm}
\caption{Relative (log) transmission of a point source in R100 (CLEAR/PRISM) mode within a 1$\times$3 NIRSpec microslit (normalised to a centred point source in the central microshutter) as a function of spatial position.  Left: the transmission for a single exposure at 2 $\mu$m and for a standard nodding pattern with 1/3 of the exposure time spent with each of the three shutters at the (0,0) position.  Right: the same but for 5 $\mu$m.  All throughputs are calculated using \texttt{Pandeia} \citep{10.1117/12.2231768}. Solid lines denote the original position of the 1$\times$3 NIRSpec microslit while the dashed lines indicate the position of a microshutter in the upper/lower nodding configurations; from top to bottom, each of the five positions observed during a full nodding sequence receives 1/3, 2/3, 1, 2/3, 1/3 of the total exposure time.  Particularly at longer wavelengths, objects that lie outside of the 1$\times$3 microslit can have a significant throughput due to the size and wings of the PSF.}
\label{fig:transmission}
\end{figure*}

Such caveats notwithstanding, we use \texttt{Pandeia} to determine limiting sensitivities for spatially- and spectrally-unresolved emission lines as a function of wavelength in each possible NIRSpec configuration.  We model the observations assuming a 1$\times$3 configuration of open microshutters in the MSA (a `microslit') and a standard three-point nodding pattern in which one third of the total exposure time is spent in each nodding position (centring on shutters $-$1, 0, and +1).  In addition, we can model the throughput of the system as a function of spatial position with respect to the open microslit.  This can be seen in  Figure \ref{fig:transmission} for the R100 prism observations as the relative throughput for single exposures for for the full three-point nodding pattern.

Significant throughput in the system occurs even when a (point source) object is not centred inside the open shutter area: an object located 0\farcs2 from the microslit centre in the dispersion direction (twice the open area) still has 10 per cent total transmission at 5 $\mu$m.  This is a direct consequence of the large wings of the JWST PSF.  Asymmetries in the PSF from the telescope and from the NIRSpec optical system lead to variations in the spatial throughput that also vary with wavelength.  This is clear from Figure \ref{fig:transmission} as a gradient in throughput along the dispersion direction at 5 $\mu$m and highlights the need for the full 3D modeling of the instrument transmission performed with \texttt{Pandeia}.  Section \ref{sec:size} further discusses the impact of the assumption of the source geometry.

In the following we assume the sources are kinematically unresolved, i.e. with velocity widths $\lesssim$ 100 km s$^{-1}$, so as not to be resolved even in the highest-resolution mode of NIRSpec ($R\sim$ 2700, otherwise known as R2700).  A spectrally-resolved line would lower the S/N estimate slightly by spreading the line flux over more pixels.  Typical $z\sim2$ `Extreme Emission Line Galaxies' have line widths of the order 50 km s$^{-1}$ \citep{2014ApJ...791...17M} which should anti-correlate with mass.  We therefore expect the majority of our serendipitous emission line sources to be spectrally unresolved in all NIRSpec observing modes.  Conversely, in R100 spectra \hbox{[O\,{\sc iii}]} $\lambda$5007 and $\lambda$4959 are not fully resolved at all wavelengths \cite[see figure 8 of][]{2017arXiv171107481C}.  As our model only considers the flux of \hbox{[O\,{\sc iii}]} $\lambda$5007 for detections, at $z\lesssim5$ we may be underestimating the true number of detectable \hbox{[O\,{\sc iii}]}-emitters.  

\subsection{Incomplete wavelength coverage with R2700}
\label{sec:r2700}

\begin{figure*}
\begin{center}
\includegraphics[width=.95\textwidth]{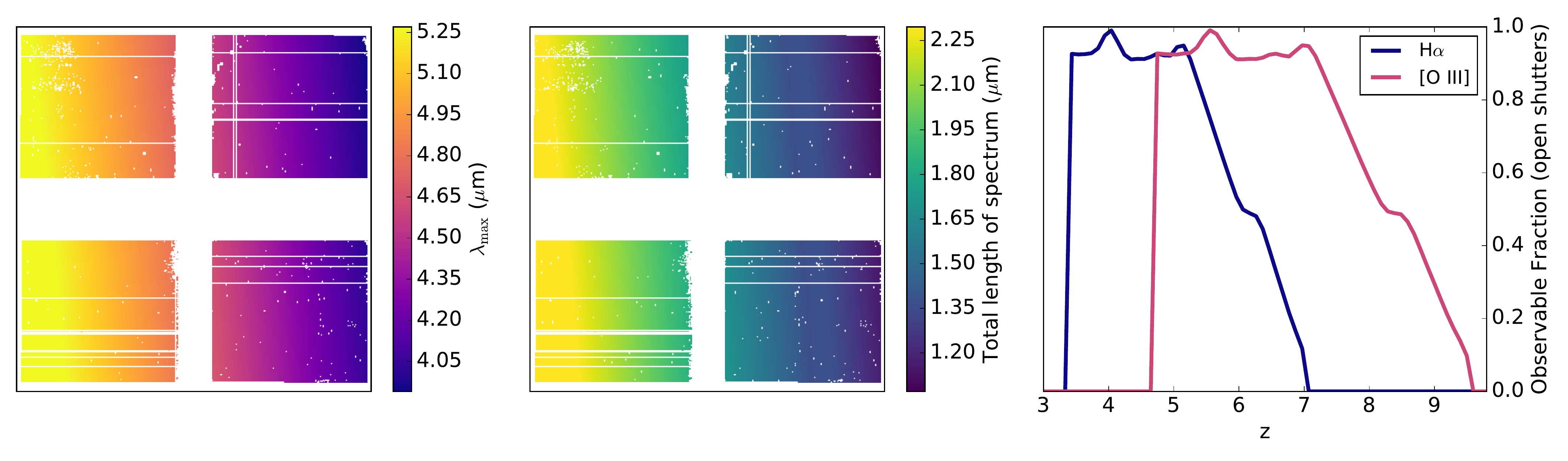}
\end{center}
\caption{Illustration of the effect of truncated spectra in the R2700 Band \hbox{{\sc iii}} mode of NIRSpec (nominal: 2.9$-$ 5.3 $\mu$m).  Left: the longest wavelength covered for a spectrum beginning at the specified MSA position, as projected onto the detector.  Middle: the total length of the R2700 Band \hbox{{\sc iii}} spectrum, taking into account truncation by the edges of the detectors; only objects in the far left of the MSA have uninterrupted wavelength coverage, excepting the detector gap. Right: total observability of {\hbox{H $\alpha$}} and \hbox{[O\,{\sc iii}]} emission in R2700 Band \hbox{{\sc iii}} as a function of redshift when combined over all open MSA shutters.  At any spatial position in the MSA, there is not an equal chance that an emission line will be observable.  The effect is similar for the other R2700 modes.}
\label{fig:r2700}
\end{figure*}

While observations with the R100 and R1000 modes can be designed to avoid the gap present between the two NIRSpec detectors, spectra obtained with the R2700 mode are too long ($\approx$3600 pixels) to avoid this gap.  By design, only spectra from MSA shutters in the $\sim$200 wide band of columns toward the extreme left edge of the MSA can yield R2700 spectra that cover the full wavelength range (modulo the detector gap).  We calculate the effect of the spectral truncation on every operable MSA shutter (193,860 in total) using the \texttt{MSAViz} code (version 1.0.4a2\footnote{\url{https://jwst-docs.stsci.edu/display/JPP/JWST+NIRSpec+MSA+Spectral+Visualization+Tool+Help}}).  Results are shown in Figure \ref{fig:r2700}.  This effect produces a clear spatial variation, with differences of nearly a factor of two in terms of total wavelength coverage between different parts of the MSA.

Because of this effect, 28 per cent of objects placed in a random microshutter in F290LP/G395H, F170LP/G235H, or F100LP/G140H, (also referred to as R2700 Band \hbox{{\sc iii}}, Band \hbox{{\sc ii}}, and Band \hbox{{\sc i}}, respectively) would lack coverage of {\hbox{H $\alpha$}} or \hbox{[O\,{\sc iii}]} even though their redshift would imply that the line would be in the nominal wavelength range covered by the configuration.  For F070LP/G140H (R2700 Band \hbox{{\sc 0}}), this number is 13 per cent.  When calculating the total number of observable serendipitous sources in each observing mode, we assume this fixed fraction of missing confirmations to average over all possible MSA configurations.

\section{Expected Numbers of Serendipitous Sources}
\label{sec:numbers}

Our fiducial model for the evolution of the UV luminosity function is used to predict a distribution of line fluxes ({\hbox{H $\alpha$}} and \hbox{[O\,{\sc iii}]}) based on the method presented in Section \ref{sec:lines}.  For a given exposure time, we calculate the sensitivity of NIRSpec as a function of wavelength for a spectrally-unresolved point source centred in a 1$\times$3 microslit.  We then scale this sensitivity as a function of spatial position according to the throughput assuming a standard three-point nodding pattern (Section \ref{sec:nirspec}).  At each spatial position we determine the fraction of the predicted line emitters that would be detectable ($>$ 5-$\sigma$ in \hbox{[O\,{\sc iii}]} and/or {\hbox{H $\alpha$}}) if the object were located at that position.  Note that this ignores (1) the potential spatial extent of the objects, as they are all assumed to be intrinsically point-like and (2) the fact that extracting the flux for an off-centre and potentially newly-discovered source will not necessarily result in the optimal signal-to-noise.  The effects of these assumptions are discussed further in Section \ref{sec:discussion}.  By taking into account the total volume probed by the NIRSpec microslits, we obtain the total number of observable line emitters per MSA configuration and hence per survey.  

Results from a R100 survey as a function of exposure time are shown in Figure \ref{fig:tintU}; objects without broadband detections in HST-based imaging (Figure \ref{fig:tintC}) or JWST-based imaging (Figure \ref{fig:tintN}) will be discussed further in the next section.  For four different exposure times (100, 50, 10, and 3 ks), we also show the distribution of redshifts and \muv values.  Longer exposure times result in a larger number of detections at \muv $\gtrsim -$16.  In addition, the largest relative increase comes at redshifts $z=3-6$.  Overall, the predicted number of detectable serendipitous sources per microslit exceeds one for an exposure time of 75 ks (21 hours).  That is to say, for every targeted source in a NIRSpec R100 observation with 21 hours of exposure time, we expect to see another source with detectable ($>$5-$\sigma$) \hbox{[O\,{\sc iii}]} and/or {\hbox{H $\alpha$}} emission \cite[cf.][]{2017AA...608A...3B}.

We can repeat this exercise for different NIRSpec observing modes, taking into account their different 2D transmission functions for their different emission line sensitivities as a function of wavelength.  These results are shown in the left panel of Figure \ref{fig:tint}.  At all integration times, the R100 prism is the most efficient at detecting serendipitous sources due to its large wavelength coverage and similar line flux sensitivity to the higher-resolution modes.  In general, the two redder settings using the $F290LP$ and $F170LP$ filters (Bands \hbox{{\sc ii}} and \hbox{{\sc iii}}) are more efficient than the two bluer settings using $F070LP$ and $F100LP$ (Bands \hbox{{\sc 0}} and \hbox{{\sc i}}) at a fixed exposure time.  This is due to the wavelength coverage of the bands and the predominance of $z\gtrsim3$ \hbox{[O\,{\sc iii}]} emitters, visible in the central panel of Figure \ref{fig:tintU}, compared to lower-$z$ emitters at long exposure times.  The effect of the incomplete wavelength coverage in the R2700 modes (Section \ref{sec:r2700}) is also visible as the nearly constant offset between the dotted and dashed curves since the line flux sensitivities for the medium- and high-resolution gratings are similar, within 7$-$8 per cent on average.

\subsection{Serendipitous objects detected in broadband imaging}
\label{sec:imaging}
When designing an MSA configuration for extragalactic spectroscopy in the early years of the mission, many JWST users will likely use CANDELS \citep{2011ApJS..197...35G,2011ApJS..197...36K} HST imaging as the source of their input photometry.  CANDELS data covers five well-known extragalactic fields: AEGIS, COSMOS, GOODS-N, GOODS-S, and UDS.  The data consists of HST WFC3 and ACS data from the optical to the near-IR.  A principle goal of CANDELS-based NIRSpec spectroscopy will be to study the properties of $z\sim2-6$ galaxies primarily via their restframe-optical emission lines.  Given the relative number density of sources in CANDELS, a user might decide to create such a census taking into account the spatial position of other sources in the field, i.e. they will preferentially choose targets with no close companions that could contribute flux inside the microslit.

\begin{figure*}
\begin{center}
\includegraphics[width=.9\textwidth]{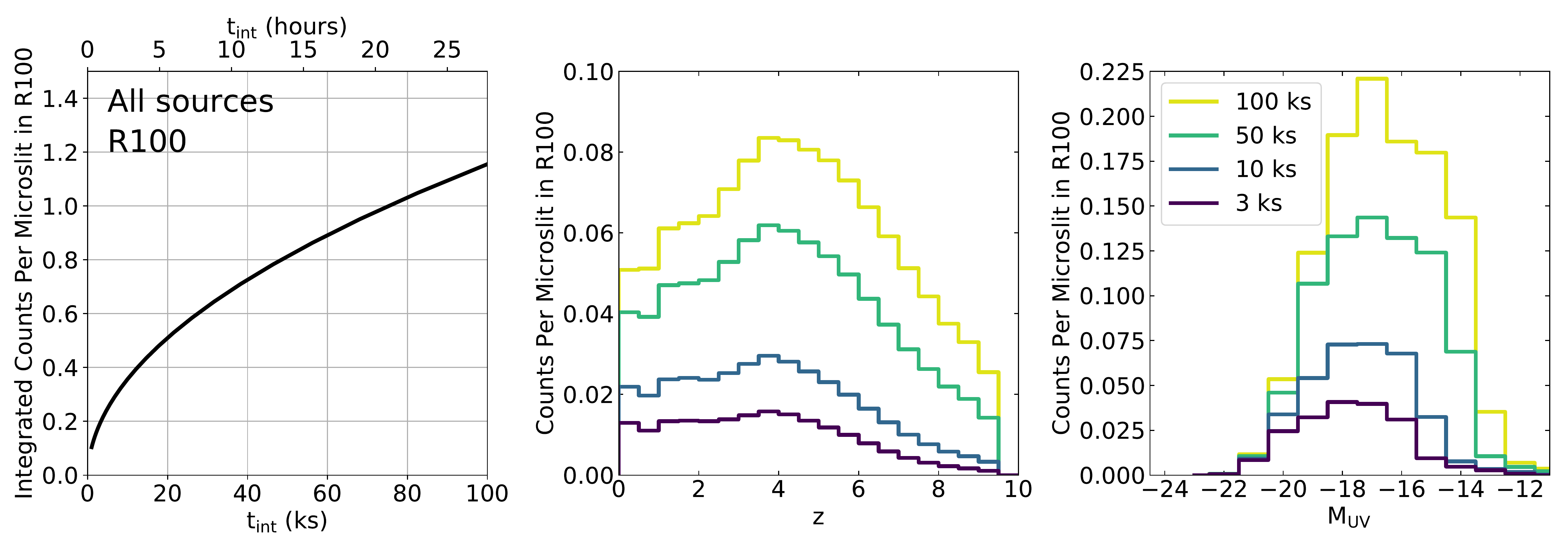}
\end{center}
\caption{(Left) Estimated cumulative counts of serendipitous emission line sources ($>$5-$\sigma$) in a R100 CLEAR/PRISM survey as a function of the exposure time.  (Middle) Redshift distributions for these sources at different exposure times.  (Right) M$_{\mathrm{UV}}$ distributions for these sources at different exposure times.  The increase in the detectable number of objects with exposure time increases approximately as a power law (deviations no more than 5 per cent) with an exponent of 0.51, implying that surveys become less efficient at detecting these sources with increasing exposure time.  The redshift and \muv distributions, though, do change with exposure time as more high-z (\hbox{[O\,{\sc iii}]}) and extremely faint (M$_{\mathrm{UV}} <$ $-$14) emitters are detectable with longer exposures.\label{fig:tintU}}


\begin{center}
\includegraphics[width=.9\textwidth]{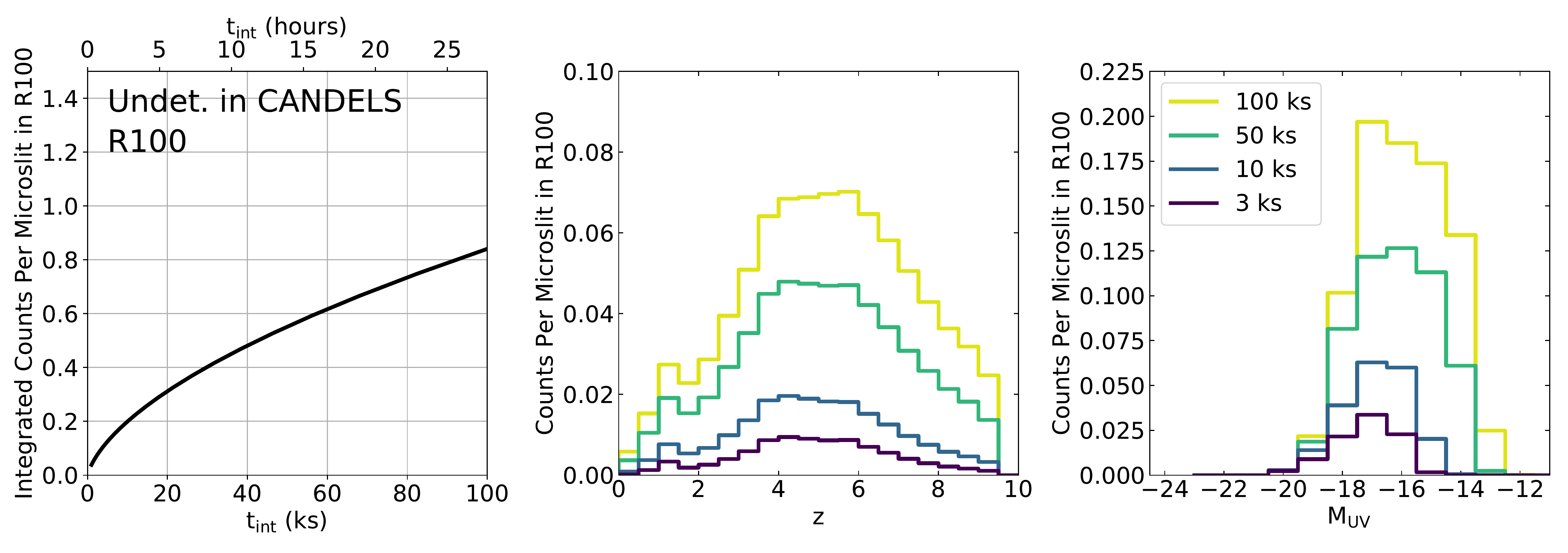}
\end{center}
\caption{Same as Figure \ref{fig:tintU} but for sources fainter than the CANDELS GOODS-S detection limits in all applicable HST bands.  At a fixed exposure time, brighter (\muv $<$ $-$19) and lower-$z$ ($z<3$) sources are less common since they are predominantly detectable in the HST imaging.\label{fig:tintC}}


\begin{center}
\includegraphics[width=.9\textwidth]{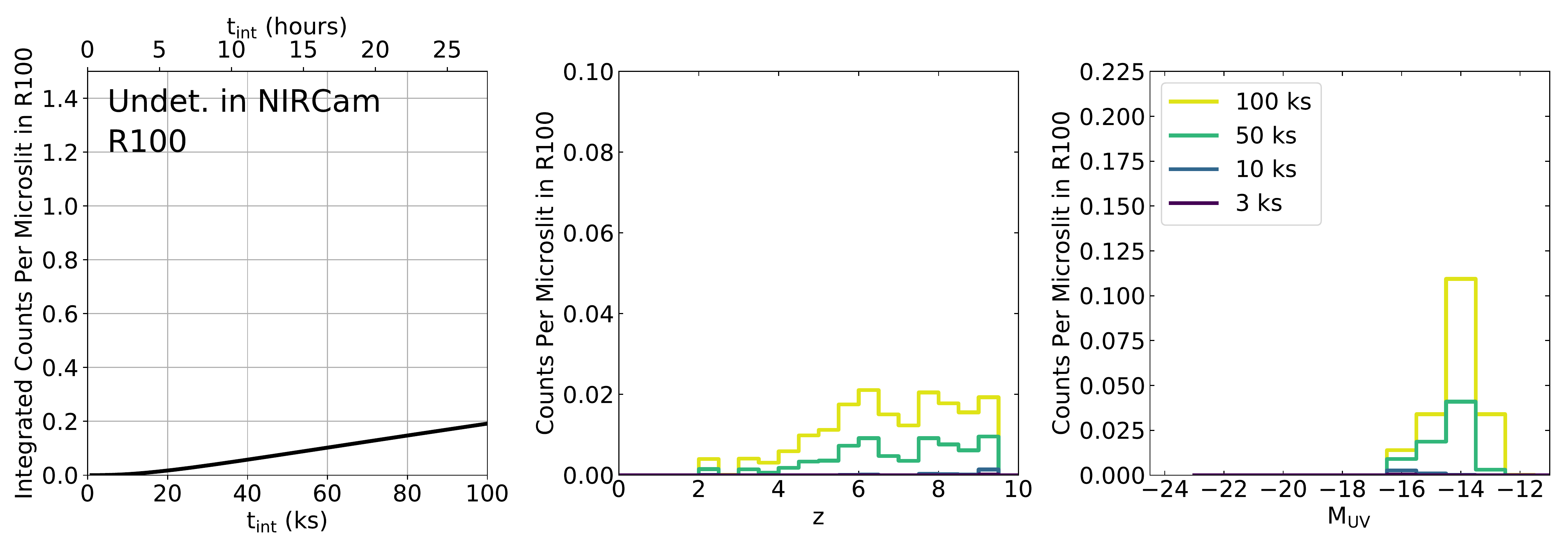}
\end{center}
\caption{Same as Figure \ref{fig:tintC} but for objects which are fainter than the NIRCam `DEEP' survey detection limits (Table \ref{tab:nc}) and the HST limits in the UDF \citep{2013ApJS..209....6I}.  Due to the depth and wavelength coverage of NIRCam, many fewer \muv $<$ $-$16 and $z<4$ galaxies remain even compared to objects that are undetected at CANDELS depth.\label{fig:tintN}}

\end{figure*}

To estimate the number of serendipitous emission line sources in this case, we follow the same procedure as above.  In addition, we apply the criterion that each galaxy must not be visible at HST wavelengths in a survey the depth of CANDELS in GOODS-S (the deepest of the five fields in many ACS and WFC3 filters) using the quoted 5-$\sigma$ values from \citet{2014ApJS..214...24S} in HST ACS/$F435W$, $F606W$, $F775W$, $F814W$, $F850LP$ and WFC3/$F125W$, $F140W$, $F160W$.  An object would be detectable if (1) its UV continuum, based on the \muv value, is bright enough to be observed directly or (2) the flux of {\hbox{H $\alpha$}}, \hbox{[O\,{\sc iii}]}, and/or {\hbox{H $\beta$}} would imply that the object is detectable in the broadband imaging.  These results are shown in Figure \ref{fig:tintC}.  

The most prominent differences with the previous case come when looking at the significantly smaller number of $z\lesssim2-3$ and \muv $\lesssim-$18 detections at all exposure times.  Filters used in the CANDELS HST imaging contain flux from \hbox{[O\,{\sc iii}]} and {\hbox{H $\alpha$}} (and {\hbox{H $\beta$}}) emission up to $z\sim2.5$, contributing to a number of detections when the EWs are sufficiently large.  In addition, the filters probe the UV continuum over the majority of the redshift range in which NIRSpec would detect the emission lines, resulting in a lower number of undetected sources with bright UV magnitudes.

Similar considerations will apply for followup of JWST/NIRCam imaging.  Table \ref{tab:nc} shows the exposure times and imaging depths achieved (for point sources) in the JWST NIRCam GTO `DEEP' program covering 46 arcmin$^2$.  Part of this imaging overlaps the Hubble Ultra Deep Field \citep[UDF;][]{2006AJ....132.1729B}, one of the most well-studied extragalactic fields, with the deepest HST imaging ever taken.  In this region, the combination of HST ACS and WFC3 imaging with NIRCam imaging will be our most complete imaging picture of the distant universe.  As above, we can determine the number of serendipitous emission line sources that would be undetectable even in the deepest 4.6 arcmin$^2$ of the UDF \citep[using the HST depths from][for the JWST/NIRCam depths from Table \ref{tab:nc}]{2013ApJS..209....6I}, with results shown in Figure \ref{fig:tintN}.  The majority of the sources detectable in exposure times of 10 ks or less are expected to have counterparts in at least one photometric band covered by the UDF and NIRCam `DEEP' surveys.  In fact, even at longer exposure times only sources with \muv $\gtrsim -$16 and $z\gtrsim4$ remain undetectable with this imaging.  The vast majority of these sources (92 per cent) are expected to be detectable via their \hbox{[O\,{\sc iii}]} emission.  As discussed in Section \ref{sec:oiii}, the assumption of a fixed \hbox{[O\,{\sc iii}]} to {\hbox{H $\alpha$}} ratio equal to 2.1 may be an overestimate in galaxies with extremely low metallicities, which are expected at the highest redshifts and faintest \muv values.  Hence, our predicted counts in this regime likely represent an upper limit.

\begin{figure*}
\begin{center}
\includegraphics[width=.99\textwidth]{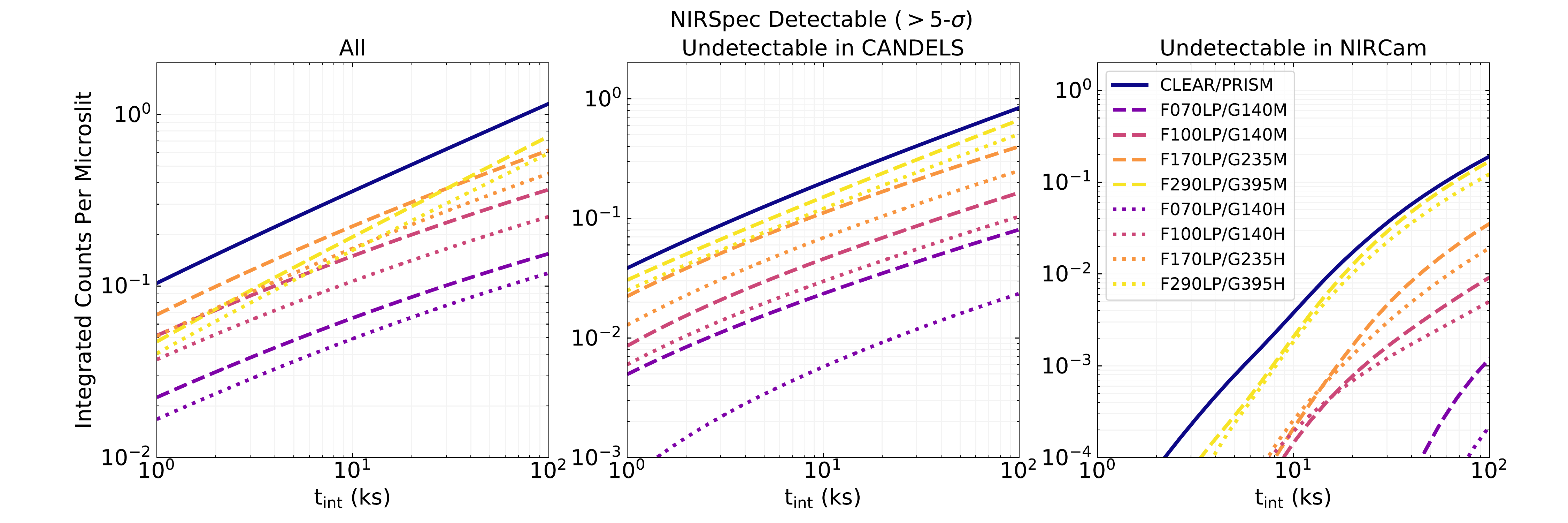}
\end{center}
\caption{Integrated counts of serendipitous emission line sources ($>$5-$\sigma$, all-$z$) per NIRSpec microslit as a function of exposure time in each grating/filter configuration.  The left panel shows the counts for all NIRSpec-detectable sources regardless of their brightness, the middle panel shows the sources that are undetectable at the depth of the CANDELS HST imaging (in GOODS-S), and the right panel shows the sources that are undetectable at the depth of a NIRCam `DEEP' survey (see Table \ref{tab:nc}).  Note the different $y$-scales between the panels.}
\label{fig:tint}
\end{figure*}

Comparisons between all three cases (all objects, objects undetectable in CANDELS imaging, and objects undetectable in the NIRCam `DEEP' survey) are shown in Figure \ref{fig:tint} for all NIRSpec configurations.  When comparing the shape of the curves, the relation between the integrated counts and the exposure time is nearly a power law for all objects for CANDELS-undetected objects.  However, the relationship is noticeably different for the NIRCam-undetected objects.  In fact, the derivative d $counts$/ d $t_{int}$ for R100 is constantly decreasing for the former two whereas it increases until $t_{int}=$ 26 ks before flattening for the latter.  This implies that the ability to detect serendipitous sources does not increase in efficiency with exposure time when considering all objects or ones that would not be detected in CANDELS-like imaging, and hence the maximal counts would be obtained by a fast tiling of the sky (the contribution of observing overheads notwithstanding).  However, for objects that would remain undetectable even in NIRCam imaging, a series of $\sim$7 hour exposures would result in the largest number of serendipitous emission line sources.

\begin{figure*}
\begin{center}
\includegraphics[width=.9\textwidth]{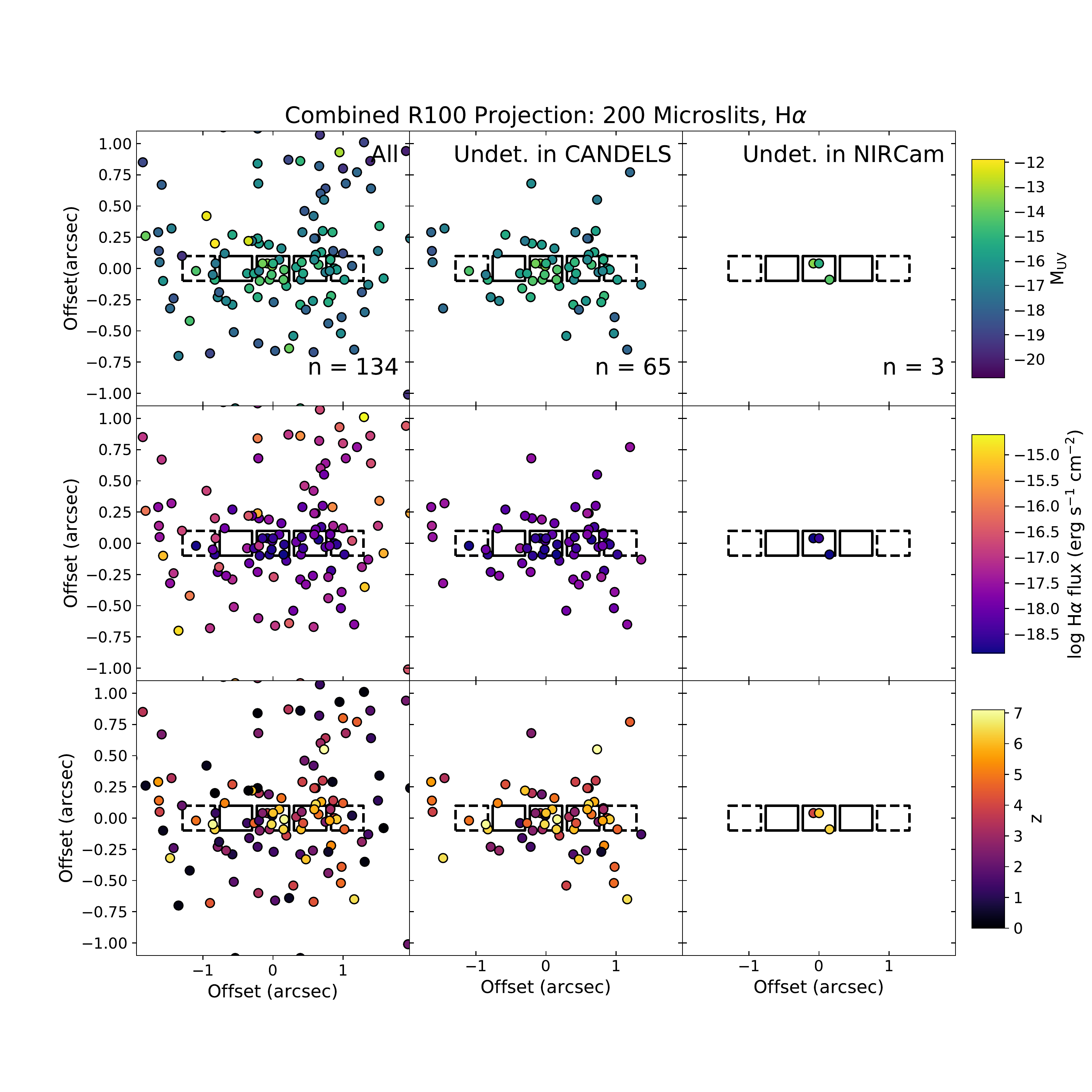}
\end{center}
\vspace{-1cm}
\caption{Combined projection of 200 NIRSpec \ot microslits (approximating a full MSA) in a 100 ks R100 survey, showing a random iteration of the spatial distribution of serendipitous {\hbox{H $\alpha$}} emitters that would be detected at $>$5-$\sigma$ with NIRSpec.  The NIRSpec microshutters are shown with the solid and dashed lines (see Figure \ref{fig:transmission}).  The sources are colour coded according to the \muv (top), {\hbox{H $\alpha$}} line flux (middle), and redshift (bottom) and split between all observable sources (left; Figure \ref{fig:tintU}), all sources that would be undetectable in CANDELS imaging (middle; Figure \ref{fig:tintC}), and all sources that would be undetectable in NIRCam `DEEP' imaging (right; Figure \ref{fig:tintN}).  The same trends as shown in Figures \ref{fig:tintU} $-$ \ref{fig:tintN} are clear here, namely the dearth of continuum-bright (\muv $<$-18) and low-$z$ ($z<2$) sources that would remain undetected in deep photometric data.}
\label{fig:shutter}
\end{figure*}

We can also create a simulated NIRSpec MSA observation by collapsing the projection of 200 individual microslits, approximating a full MSA, along the wavelength direction and combining them into a single 2D map of serendipitous detections, assuming that the sources are distributed randomly on the sky.  Figure \ref{fig:shutter} illustrates this collapsed image for serendipitous {\hbox{H $\alpha$}} emitters in the three cases outlined above, with colour-coding in each panel corresponding to $M_{\mathrm{UV}}$, {\hbox{H $\alpha$}} flux, and $z$.  The same trends as in Figures \ref{fig:tintU} $-$ \ref{fig:tintN} manifest themselves in these projections, namely a significantly smaller number of $z\lesssim4$ and \muv $\lesssim-$18 sources in the samples of CANDELS- and NIRCam-undetected emitters.  The spatial distribution of the sources also changes due to these imaging requirements as the brightest sources with $f_{\mathrm{H\alpha}} > 10^{-16}$ \cgs, detectable far from the open microslit, would predominantly appear in the broadband imaging.  Although our analysis does not consider the potential effect of spatial clustering in the galaxy population, clustering strength decreases with $M_{\mathrm{UV}}$ \citep{2005ApJ...619..697A} and the counts are dominated by faint galaxies.  

Considering the small angular separations between some serendipitous sources and the centre of the microslit (Figure \ref{fig:shutter}), it is tantalizing to think that magnification from strong gravitational lensing might allow for the detection of even fainter sources and hence further increase the expected counts.  In practice, the incidence rate is difficult to establish as the lensing magnification factor at a fixed angular separation depends on the mass and mass profile of the lensing source, i.e. the primary target of the NIRSpec observations, which are not explicitly considered here.  Massive, elliptical galaxies are thought to dominate the total lensing probability of the universe \citep{1984ApJ...284....1T}, so observations targeting these galaxies might result in more detections of lensed background sources.  Indeed, two similar instances have been discovered in deep CANDELS HST imaging where an elliptical galaxy at $z=0.65$ strongly lenses an `Extreme Emission Line Galaxy' at $z=1.85$ \citep{2012ApJ...758L..17B} and similarly with a lensing elliptical galaxy at $z=1.53$ and an EELG at $z=3.42$ \citep{2013ApJ...777L..17V}.  These observations also highlight the number density of $z>1$ galaxies with high-EW optical emission lines.

\begin{table*}
\begin{center}
\begin{tabular}{|l|cccc|ccccc|}
\multicolumn{10}{c|}{NIRCam `DEEP' GTO Imaging} \\
\hline
 &  F090W &  F115W &  F150W &  F200W &  F277W &  F335M &  F356W &  F410M &  F444W \\
\hline
Exposure time (ks)      & 60.5 & 80.8 & 59.3 & 38.8 & 49.1 & 30.9 & 38.8 & 60.5 & 60.5 \\
5-$\sigma$ Point Source Magnitude (AB) 	     & 30.3 & 30.6 & 30.7 & 30.7 & 30.3 & 29.6 & 30.2 & 29.8 & 29.9 \\
\hline
\end{tabular}
\end{center}
\caption{\label{tab:nc}Summary of NIRCam Imaging Program (Proposal ID 1180; PI: D. Eisenstein) covering 46 arcmin$^2$.  
({\it top}) Average exposure times per filter.  As pointings often overlap, significantly deeper (2$\times$) regions will exist in the final mosaics.
({\it bottom}) The 5-$\sigma$ depth for a point source corresponding to the
average exposure times.  See also \citet{2018ApJS..236...33W}.
}
\end{table*}

\section{Discussion}
\label{sec:discussion}

\subsection{Comparison to  \hbox{[O\,{\sc iii}]} in the 3D-HST survey}
\label{sec:3dhst}

The large spatial area covered by open microslits in a NIRSpec MSA survey is in many ways similar to a large-field slitless spectroscopic survey.  In fact, we can use statistics about the frequency of emission lines in existing slitless (grism) spectroscopic surveys to test our model of emission line fluxes and equivalent widths.  The 3D-HST survey \citep{2012ApJS..200...13B,2016ApJS..225...27M}, which uses the G141 grism on the HST/Wide Field Camera 3 (WFC3), provides near-complete (modulo contamination) spectral coverage over 626.1 arcmin$^2$ from 1.1 to 1.65 microns with a spectral resolution $R\sim130$.

To compare with 3D-HST, we assume that all emission lines are spectrally- and spatially-unresolved \cite[the latter is explicitly taken into account when determining the limiting sensitivity of the survey;][]{2016ApJS..225...27M}.  Down to m$_{F140W} =$ 26.0, 3D-HST has $>$5-$\sigma$ detections of 4972 \hbox{[O\,{\sc iii}]} emitters at $1.1 < z < 2.6$ \citep{2016ApJS..225...27M}.  Using our model and enforcing the same $F140W$ magnitude limit, we would expect 4474 \hbox{[O\,{\sc iii}]} emitters at the same redshifts over the same spatial area, or $\approx$ 10 per cent less than in the full 3D-HST catalogue.  We note that this includes objects with low-EW emission lines which are unlikely to be observable at the highest redshifts.

Given that the background noise level in the grism exposures varies by nearly a factor of 3 across the five 3D-HST fields \citep{2012ApJS..200...13B}, and that the derived emission line sensitivity is only an average measurement, we conclude that our emission line and equivalent width model is compatible with 3D-HST observations of $1.1 < z < 2.6$ \hbox{[O\,{\sc iii}]} emitters.  These data are a close approximation of the JWST/NIRSpec regime at low-$z$.

\subsection{Evolution in the model EW with redshift and \muv}

In our model for the evolution of {\hbox{H $\alpha$}} EW with $z$ and \muv (Equation \ref{eqn:ew}), we use the redshift evolution from \citet{2013ApJ...777L..19L}, EW$_{\mathrm{H\alpha}} \propto$ (1$+z$)$^{1.2}$.  This power-law slope implies a factor of $\sim$2 slower evolution than that found in \citet{2012ApJ...757L..22F} at $z < 2$, EW$_{\mathrm{H\alpha}} \propto$ (1$+z$)$^{1.8}$.  Similarly, \citet{2016MNRAS.460.3587M} find an even slower evolution, EW$_{\mathrm{H\alpha}} \propto$ (1$+z$)$^{1.0}$.  A different evolution with redshift would alter the predictions for the expected number of NIRSpec-detectable serendipitous sources.  In particular, we would observe a different number of \hbox{[O\,{\sc iii}]} emitters at $z\gtrsim4-5$, where the observed counts in Figure \ref{fig:tintU} begin to turn over.  In 100 ks at R100, EW$_{\mathrm{H\alpha}} \propto$ (1$+z$)$^{1.2}$ predicts 1.23 detectable ($>$5-$\sigma$) sources per microslit, 0.93 of which would be undetectable at the depth of CANDELS imaging and 0.23 at the depth of the `DEEP' NIRCam imaging;  EW$_{\mathrm{H\alpha}} \propto$ (1$+z$)$^{1.0}$ predicts 1.14, 0.82, and 0.17 sources per microslit, respectively (cf. 1.15, 0.84, and 0.19 using our fiducial model).

Only the steeper EW evolution with redshift predicts counts that differ by more than 10 per cent.  With this evolution, we would expect a larger number of faint objects (based on the UV luminosity function) to have high-EW emission lines and satisfy these criteria: the difference in EWs between the \citet{2012ApJ...757L..22F} evolution and the \citet{2013ApJ...777L..19L} evolution is a factor of 3.7 at $z=8$.  As stated previously, the fact that \citet{2013ApJ...777L..19L} base their evolution on measurements that span from $z\sim1-8$ \citep[or $z=1-5$ for][]{2016MNRAS.460.3587M} compared with just $z\sim0.6-1.6$ makes it more applicable to high-$z$ studies with NIRSpec.  

Similarly, we assume the slope $d~\mathrm{log_{10}EW_{H\alpha}}/d~M_{\mathrm{UV}}$=0.08 from \citet{2016ApJ...833..254S}.  This relationship is measured from galaxies with $-$22 $<$ \muv $<$ $-$20, much brighter than the galaxies that we predict dominate the serendipitous number counts.  If we were instead to assume a slope of 0.21 \cite[the 1-$\sigma$ upper limit from ][]{2016ApJ...833..254S}, we would expect 2.0 times more serendipitous counts at a fixed $t_{int}$.  If there were no relationship (i.e. a slope of 0), then the model predicts 0.6 times the number of serendipitous counts.  Thus over a broad range in $d~\mathrm{log_{10}EW_{H\alpha}}/d~M_{\mathrm{UV}}$, the predicted serendipitous counts can vary by a total factor of 3.1.  This is driven by differences at the faint end as the predicted counts at \muv $>$ $-$18 vary by less than 10 per cent.

The exact slope and scatter of these relations will be robustly constrained in the first large spectroscopic surveys with NIRSpec, as well as large imaging surveys using NIRCam and MIRI.

\subsection{Star formation histories\label{sec:sfh}}

In the conversion of {\hbox{H $\alpha$}} EW to {\hbox{H $\alpha$}} flux, we could assume a continuous star formation history as opposed to an instantaneous burst.  \texttt{Starburst99} can reproduce the typical {\hbox{H $\alpha$}} EWs predicted by our parameterization for a continuous star formation episode of 10$-$100 Myr, similar to the length of star formation episodes in present-day low-mass galaxies \cite[$M_{\star} <$ 10$^{7.5}$ \msol;][]{2018arXiv180906380E} and those predicted by many simulations of galaxy formation at $z > 6$ \cite[e.g.][]{2018MNRAS.480.4842C}.  As shown in the lower panel of Figure \ref{fig:ew}, this star formation history results in higher predicted {\hbox{H $\alpha$}} luminosities at a fixed EW since the underlying stellar population is older and hence has a brighter optical continuum.

These higher luminosities lead to higher line fluxes and hence a larger number of predicted serendipitous counts.  Compared to our fiducial model, a model using continuous star formation predicts 30 per cent more counts at a fixed $t_{int}$.  These additional counts are predominantly at $z<6$ (75 per cent of the excess counts) where the {\hbox{H $\alpha$}} EWs are lower overall and hence the age difference between the two different star formation histories is largest.

\subsection{Physical extent of the serendipitous emitters\label{sec:size}}

In our fiducial model we treat all objects as point-like: a larger source centred inside the microslit would have a lower S/N due to additional slit losses, but non-centred sources would also potentially have more of their flux falling into the microslit.  In order to quantify these effects, we consider the case of extended sources with  S\'ersic indices $n=4$ and $n=1$ and half-light radii of 0$\farcs$2, which is the average projected size of $z=3-6$ galaxies \citep{2015ApJS..219...15S,2016MNRAS.457..440C}.

If we adopt a fixed 0$\farcs$2 size and S\'ersic indices of $n=1$ or $n=4$ instead of a point source geometry for all objects, we predict a larger number of serendipitous counts at a fixed exposure time.  The largest number of serendipitous counts is predicted for the $n=4$ S\'ersic profile which, for a fixed half-light radius, has the most flux at radii $>$ 1$''$.  In general, extended sources have their flux distributed over a larger area and, as most serendipitous counts come from objects that are not located inside the open area of the microslit (Figure \ref{fig:shutter}), this results in significantly more detectable emitters at $z < 6$ and \muv $<-$14.  The total counts differ by a factor of 1.64 ($n=4$) and 1.21 ($n=1$) at 100 ks compared to our fiducial model, but this drops to a factor of 1.34 and 1.16, respectively, when restricting to sources with $z>6$ (which are also more likely to be spatially-compact).  When restricting to sources with \muv $>-$14.5, the $n=4$ and $n=1$ counts are lower than the fiducial model by factors of 0.96 and 0.91, respectively, since these faint sources need to be well-centred and a point source geometry results in a higher flux throughput.


Current data, however, indicates that clumpy star formation is more prevalent at high redshifts \cite[e.g.][]{2015ApJ...800...39G}.  These clumps do not necessarily have the same emission line properties as the main stellar body of the galaxy \cite[e.g.][]{2015Natur.521...54Z} and their surface brightnesses are observed to increase with redshift \citep{2015MNRAS.450.1812L}.  Therefore a point-source assumption would be appropriate for objects if there is on-average one luminous star-forming clump per galaxy that dominates the line emission of the galaxy; deep JWST/NIRCam imaging probing both the restframe-UV and restframe-optical will shed more light on this issue at the relevant redshifts.  Extended sources would also add complications to the detection of these sources as the flux would likely be spread over more detector pixels.  Standard methods for background subtraction become more complex with an additional source of flux within a microslit, so observers would need to consider alternatives such as averaging the background level in multiple dedicated sky microslits.

\subsection{Comparison to line fluxes in JAGUAR}
\label{sec:jaguar}

\citet{2017arXiv171107481C} use \texttt{BEAGLE} \citep{2016MNRAS.462.1415C} to make predictions for JWST observations based on existing photometric samples in the UDF, modeling and predicting the radiation coming from realistic stellar populations and the effects on the gas within the galaxy.  \citet{2018ApJS..236...33W} take this analysis further by creating a full `mock' catalogue (`JAGUAR') of sources based on existing mass and luminosity functions, with extrapolations to higher redshifts, lower masses, and lower luminosities than are currently constrained with observations.  This includes information about the broadband SEDs of the galaxies for predictions for the fluxes of various emission lines.

\begin{figure}
\begin{center}
\includegraphics[width=.45\textwidth]{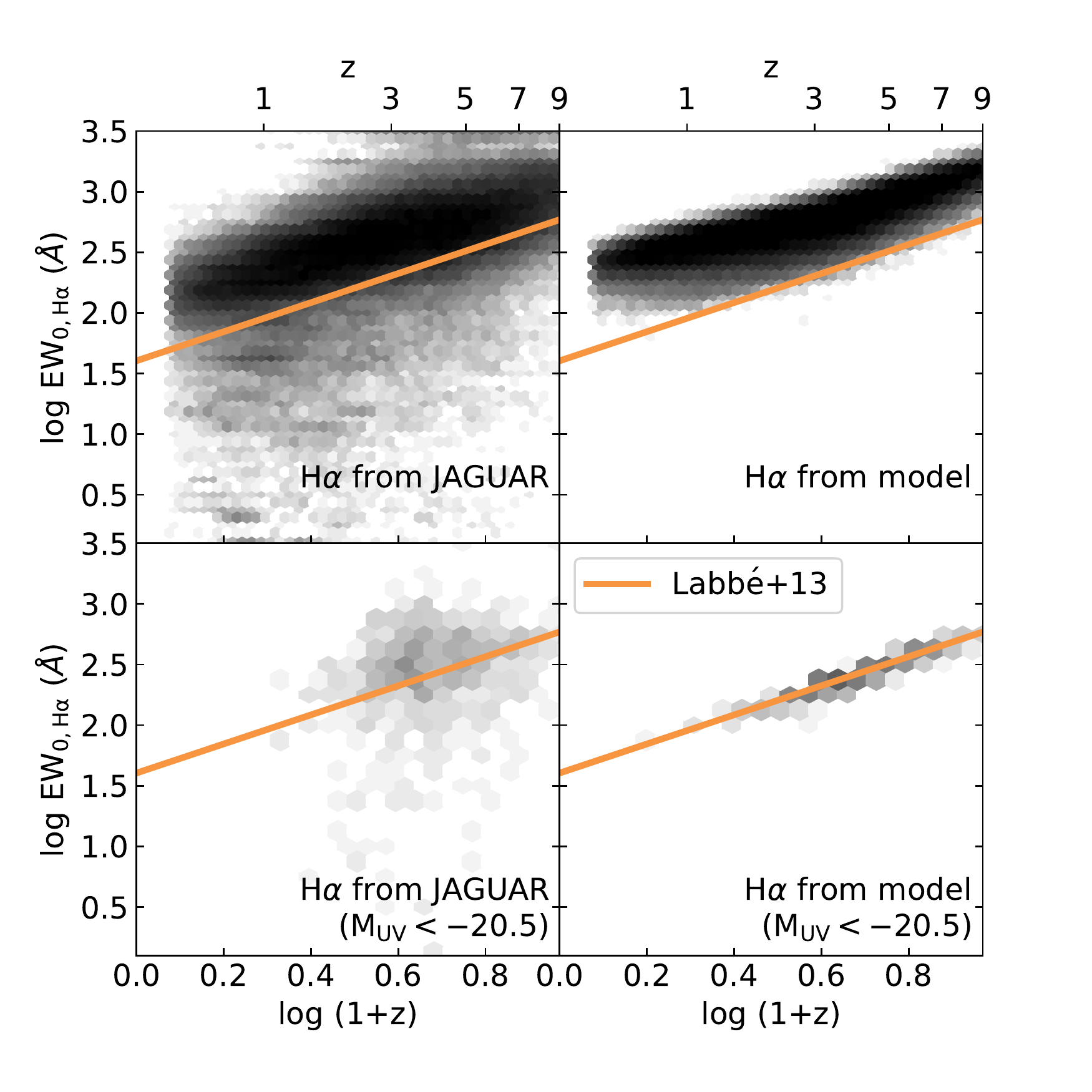}
\end{center}
\vspace{-0.5cm}
\caption{{\hbox{H $\alpha$}} EW versus redshift for the full \citet{2018ApJS..236...33W} JAGUAR catalogue (top-left) and from the prediction based on \muv (Equation \ref{eqn:ew}, top-right) with (log) density in each bin shown with shading.  The relationship from \citet{2013ApJ...777L..19L} is overplotted as the solid line and appears to under-predict the {\hbox{H $\alpha$}} EWs at all redshifts.  However, as noted in Section \ref{sec:lines}, there is a dependence on $M_{\mathrm{UV}}$: \citet{2013ApJ...777L..19L} only measure EWs in galaxies above 1.5 M$^*$ at $z=8$, corresponding to \muv $= -$20.5.  When we restrict the JAGUAR catalogue to values brighter than this, we obtain a better agreement (bottom panels).}
\label{fig:jaguar2}
\end{figure}

Figure \ref{fig:jaguar2} compares our method for obtaining {\hbox{H $\alpha$}} EWs from the UV luminosity (right panels) to the JAGUAR EWs from \citet[][left panels]{2018ApJS..236...33W}.  The mean ratio of the JAGUAR {\hbox{H $\alpha$}} EW and the $M_{\mathrm{UV}}$-derived EW is 0.78 for galaxies of all magnitudes.  As can be seen, the distribution of EWs does not match the relationship between EW and redshift derived in \citet{2013ApJ...777L..19L}.  However, when restricting to the same \muv range as \citet[][i.e. \muv $< -$20.5]{2013ApJ...777L..19L}, then the distribution of both JAGUAR and $M_{\mathrm{UV}}$-derived EWs with redshift matches well (bottom panels).  In particular, the agreement in the bottom-right panel highlights the agreement between the observations of \citet{2013ApJ...777L..19L} and those of \citet{2016ApJ...833..254S}, both of which are used to derive Equation \ref{eqn:ew} and both of which rely on photometric excesses in IRAC data to measure optical line EWs at $z>4$.

\citet{2018ApJS..236...33W} use this catalogue to predict the number counts of high-$z$ galaxies in NIRCam surveys, but it can also be used to predict the number of serendipitous sources that would be detectable with NIRSpec.  To determine the number of serendipitous sources from this catalogue, we place a series of 10,000 microslits in random positions within the JAGUAR catalogue field-of-view.  We then apply the NIRSpec throughputs (Figure \ref{fig:transmission}) to mimic an observation and nodding sequence and determine how many objects would have their {\hbox{H $\alpha$}} and/or \hbox{[O\,{\sc iii}]} emission detected as a function of the total exposure time.  The catalogue also contains HST and JWST broadband fluxes for each object (including the contribution of the emission lines) so CANDELS and NIRSpec `DEEP' detectability are, as before, defined to be any broadband magnitude greater than the nominal 5-$\sigma$ survey limit in that band.

\begin{figure}
\begin{center}
\includegraphics[width=.45\textwidth]{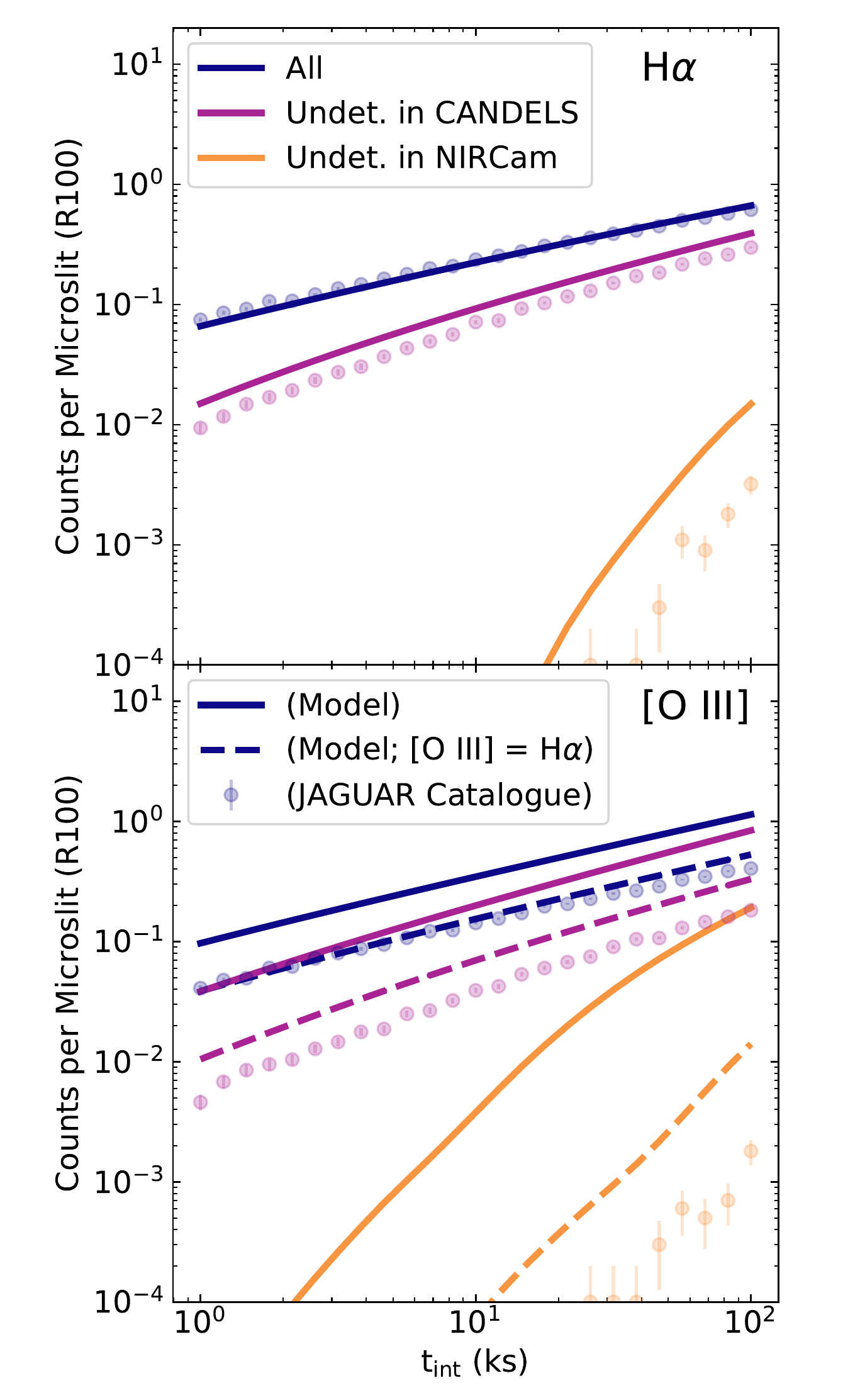}
\end{center}
\caption{Comparison of the estimated counts of NIRSpec-detectable ($>$5-$\sigma$) serendipitous sources from our model (solid lines) and from simulated observations from the JAGUAR catalogue of \citet[][poisson counting errors included]{2018ApJS..236...33W}; the top panel is for {\hbox{H $\alpha$}} emitters and the bottom panel is for \hbox{[O\,{\sc iii}]} emitters.  Colours differentiate the three different regimes: all galaxies (blue), sources undetectable in CANDELS imaging (purple), and sources undetectable in NIRCam `DEEP' imaging (orange).  The counts for \hbox{[O\,{\sc iii}]} are noticeably higher in the model than in the JAGUAR catalogue as we assume a higher average value for \hbox{[O\,{\sc iii}]}/{\hbox{H $\alpha$}}, but when setting the fluxes of the lines to be the same the agreement is much better (dashed lines, bottom panel).}
\label{fig:jaguar}
\end{figure}

Results for R100 are illustrated in Figure \ref{fig:jaguar}, also including the model predictions from Figures \ref{fig:tintU} $-$ \ref{fig:tintN}.  The predicted {\hbox{H $\alpha$}} counts (top panel) are comparable to within a factor of $\times$1.4 for the total number as well as for the CANDELS-undetected objects and within a factor of $\times$5 for NIRCam-undetected objects.  The predictions for \hbox{[O\,{\sc iii}]} counts differ by a larger amount: a factor of $\times$2.4 and $\times$5.3 for the total number and the CANDELS-undetected objects and a factor of $\times$200 for the NIRCam-undetected objects (solid lines; lower panel).  If we were to assume a lower EW for \hbox{[O\,{\sc iii}]}, specifically setting the flux to be equal to that of {\hbox{H $\alpha$}}, we obtain the dashed curves which agree with the JAGUAR catalogue results similarly to the case of {\hbox{H $\alpha$}}.

In general, the predictions from the JAGUAR catalogue and from our $M_{\mathrm{UV}}$-based model agree well for {\hbox{H $\alpha$}} and we can understand the discrepancy for \hbox{[O\,{\sc iii}]} due to differences in the relative strength of that emission line.  We will discuss the strength of \hbox{[O\,{\sc iii}]} in the next section.  Nevertheless, this comparison shows that we do expect significant numbers of serendipitous emission line sources in deep NIRSpec observations, even when considering {\hbox{H $\alpha$}} alone.  The slopes of the relations (d \textit{counts} / d $t_{int}$) from the model and from JAGUAR are strikingly similar even when the normalisations are different.  This similarity further reinforces our model relating emission line strength to the UV continuum luminosity functions, which rise as a power-law at the faint end where the majority of the serendipitous sources lie.

\subsection{Dependence on the \hbox{[O\,{\sc iii}]} to {\hbox{H $\alpha$}} ratio}
\label{sec:oiii}

As mentioned in Section \ref{sec:lines}, we adopt a median ratio of \hbox{[O\,{\sc iii}]} to {\hbox{H $\alpha$}} of 2.1  This is slightly higher than what is assumed in e.g. \citet{2003AA...401..1063A}, which results from our use of observational constraints from emission line-dominated galaxies at $z\sim2$ as well as different stellar population models. Specifically, we use the models of \citet{2016MNRAS.462.1757G} at $Z=0.004$ which were specifically designed to reproduce nebular emission in galaxies that span a broad range in physical properties and redshifts.  These models do not completely rely on calibrations from \hbox{H{\sc ii}} regions and local galaxies which do not necessarily represent the range in physical conditions expected in the average galaxy at high redshift \citep[e.g.][]{2008MNRAS.385..769B,2010ApJ...719.1168E,2014MNRAS.445.3200S,2015ApJ...801...88S,2016ApJ...816...23S,2017ApJ...836..164S,2018MNRAS.479.3264C}.  Dust is not explicitly included in our model, which would also lower the observed \hbox{[O\,{\sc iii}]}/{\hbox{H $\alpha$}} ratio from the intrinsic ratio.  However, dust attenuation is not expected to contribute strongly at low masses \citep{2010MNRAS.409..421G, 2010AA...515A..73S} and specifically for $z\gtrsim1$ galaxies with high-EW optical emission lines \citep{2014ApJ...791...17M}.

\begin{figure}
\begin{center}
\includegraphics[width=.45\textwidth]{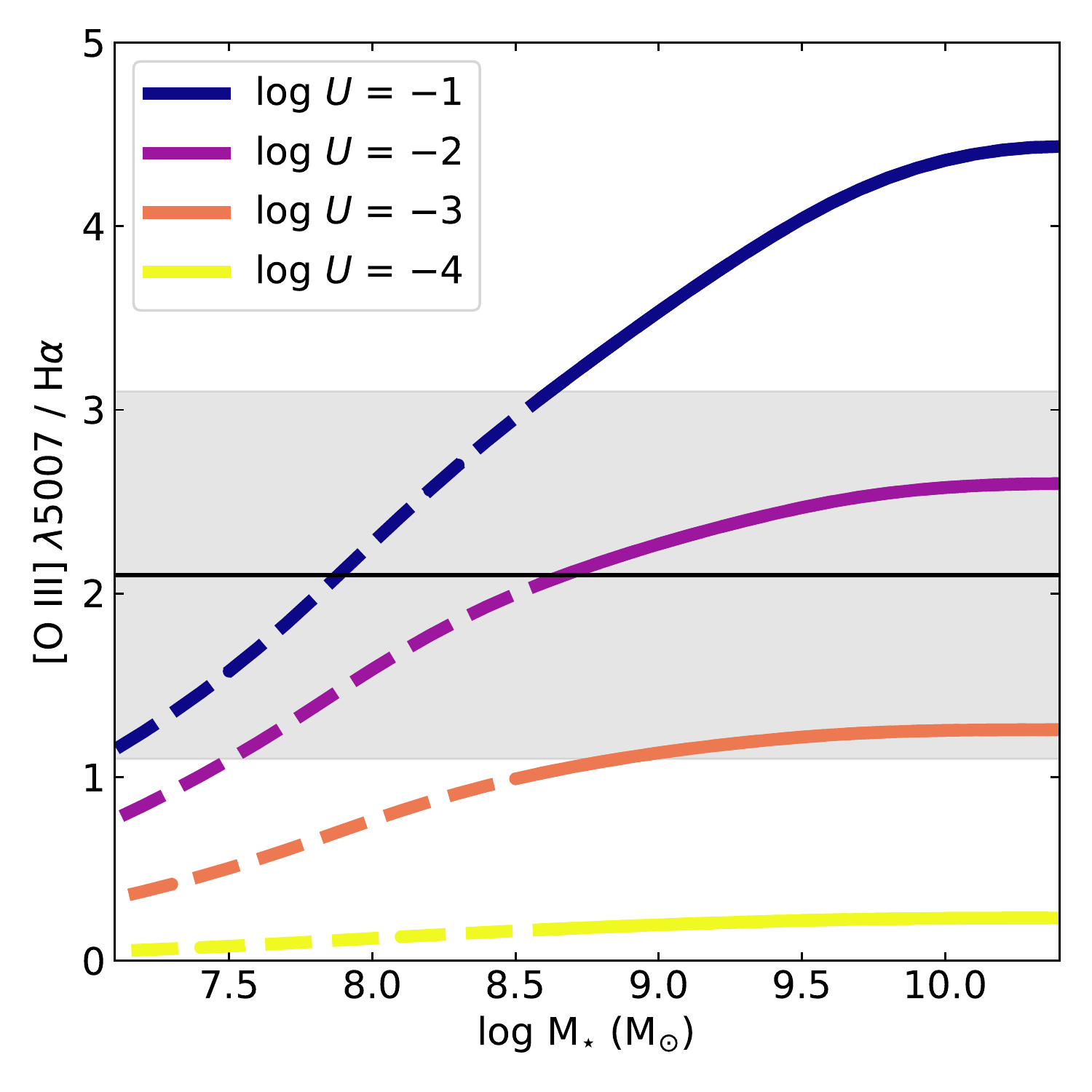}
\end{center}
\vspace{-0.5cm}
\caption{Ratio of \hbox{[O\,{\sc iii}]} $\lambda$5007 to {\hbox{H $\alpha$}} versus stellar mass from the \citet{2016MNRAS.462.1757G} for different log $U$ values and assuming the mass-metallicity relation from \citet{2010ApJ...715L.128A}, with the shaded region showing our 1-$\sigma$ adopted value of 2.1 $\pm$ 1.  The chosen mass metallicity relation is measured for `Extreme Emission Line Galaxies' at $z\sim0.3$, with dashed portions of the lines representing extrapolations to lower masses.  Galaxies with intense radiation fields \citep[log $U \gtrsim -2$;][]{2014ApJ...788L...4A} are expected to contribute the most to the serendipitous counts, although at extremely low masses and metallicities the number of \hbox{[O\,{\sc iii}]} detections could begin to drop.}
\label{fig:oiiiha}
\end{figure}

As a result of using this ratio, we predict more detectable \hbox{[O\,{\sc iii}]} emitters than the \citet{2018ApJS..236...33W} JAGUAR catalogue described in the previous section (when using \hbox{[O\,{\sc iii}]}/{\hbox{H $\alpha$}} $=$ 1, we obtain the dashed curves in Figure \ref{fig:jaguar}, which provides better agreement with the JAGUAR catalogue).  This ratio changes with galaxy properties such as metallicity and ionization parameter \citep{2018ApJ...859...84H} which in turn evolve with redshift, and hence our fixed ratio represents a simplification of the true emission line properties of high-$z$ galaxies.  This ratio is predicted to drop at the lowest metallicities (and hence stellar masses), but the mass at which this happens depends on the exact mass-metallicity relationship assumed and this regime is not yet observed. Figure \ref{fig:oiiiha} shows how this ratio evolves with stellar mass (via the mass-metallicity relationship) and log $U$ for the \citet{2016MNRAS.462.1757G} models assuming $\xi _d=0.3$, $n_H=100$ cm$^{-3}$, C/O = C/O$_{\odot}$, and $m_{up}=100$ M$_{\odot}$.  While the $z\sim0.3$ mass-metallicity relationship for `Extreme Emission Line Galaxies'  \citep{2010ApJ...715L.128A} is the closest proxy for the types of galaxies we expect to be producing strong emission lines at high-$z$; even assuming the relation from \citet{2014AA...563A..58T} at $z\approx3.5$  for all star-forming galaxies produces similar evolution in the ratio at masses $>$ 10$^{7.5}$ M$_{\odot}$.  Observations of these emission line-dominated galaxies imply log $U \approx-2$ to $-1$ \citep{2010ApJ...719.1168E,2014ApJ...788L...4A,2018ApJ...859..164B} which is much larger than the values for local star-forming galaxies used in \citet{2018ApJS..236...33W} based on the observations of \citet{2017MNRAS.468.2140C}.  This is also true for the general population of \hbox{[O\,{\sc iii}]}-emitting galaxies at $z>3$, which have higher ionization parameters at fixed stellar mass or metallicity than locally \citep{2017ApJ...849...39S}.

The true distribution of this ratio will not be known until JWST/NIRSpec assembles large samples of high-$z$ galaxies with detections of both lines.  Our simple model uses a ratio based on observations of \hbox{[O\,{\sc iii}]} and {\hbox{H $\alpha$}} in high-EW (high specific star formation rate), low-metallicity ($<$ 25 per cent Z$_{\odot}$), 10$^{8-9}$ M$_{\odot}$ galaxies at $z\sim1-2.5$ \citep{2018ApJ...854...29M} which have similar \hbox{[O\,{\sc iii}]} and {\hbox{H $\alpha$}} equivalent widths to the galaxies that we expect to be serendipitously detectable with NIRSpec.  While the agreement between the predictions and the 3D-HST observations described in Section \ref{sec:3dhst} gives observational evidence at $z\approx2$ for our assumption about the ratio of \hbox{[O\,{\sc iii}]} to {\hbox{H $\alpha$}}, the relationship between emission line-dominated objects and the general population of star-forming galaxies at higher redshifts remains to be seen.  In the most pessimistic case where \hbox{[O\,{\sc iii}]} is never detected because it is intrinsically much fainter than {\hbox{H $\alpha$}}, our predicted counts would be given by the curves in the upper panel of Figure \ref{fig:jaguar} where we would still achieve 0.67 counts per open microslit in 100 ks.

\subsection{Serendipitous versus targeted counts at high-$z$}

Given the expected number of serendipitous $z \gtrsim 6$ galaxies observable in a NIRSpec survey (see Figure \ref{fig:tintU}), it is natural to ask if the overall number counts of such high-$z$ galaxies will be dominated by serendipitous discoveries or by targeted spectroscopy. This high-redshift frontier represents one of the key aspects of the JWST mission in general and specifically one of the drivers for approved GTO and ERS programs.

As mentioned above, \citet{2015ApJ...803...34B} detect 137 $z>6$ photometric dropout candidates in the deep-WFC3 area UDF (123 $\times$ 136 arcsec).  This implies a source density of 29.5 objects per arcminute$^2$.  According to \citet{overbooking}, the average maximum number of non-overlapping R100 spectra at this source density is 65.  In practice, this is an overestimate given that the objects are distributed over an area much smaller than the full NIRSpec field-of-view.  Nevertheless, 65 $z>6$ spectra is still less than the 78 (0.39 per microslit) we would expect to serendipitously detect in a deep MSA survey with 200 open microslits at 100 ks.

The 11 $\times$ 11 arcminute JWST JAGUAR catalogue of \citet{2018ApJS..236...33W} contains 9355 $z>6$ galaxies with UV magnitudes that would make them detectable in the DEEP component of the NIRCam/NIRSpec joint GTO program (see Table \ref{tab:nc}), implying a source density of 77 objects per arcminute$^2$ or 110 non-overlapping R100 spectra per MSA configuration \citep{overbooking}.  In the area of the deepest planned NIRSpec observations (Proposal ID 1210; PI: P. Ferruit) of 100 ks, covering the UDF as followup of the NIRCam DEEP observations, we would therefore expect similar numbers of serendipitous $z>6$ and targeted detections.

\subsection{Effect of the faint end of the UV luminosity function}
\label{sec:nonschechter}
Given the discussion in the literature about the faint-end shape of the UV luminosity function at $z\sim6$, we consider the effect that such differences would have on the observable number of serendipitous emission line sources.  To do so we perform calculations as before, but instead of our fiducial (Schechter) luminosity function we use the \citet{2018MNRAS.479.5184A} and \citet{2017ApJ...843..129B} luminosity functions, which include a departure from a power-law at \muv $< -$16, and the Schechter function fits from \citet{2017ApJ...835..113L} and \citet{2018ApJ...854...73I}.  The last two power-law luminosity functions produce observable counts that are similar within 5 per cent, so for the purposes of this work we consider them as a single luminosity function.

\begin{figure}
\begin{center}
\includegraphics[width=.45\textwidth]{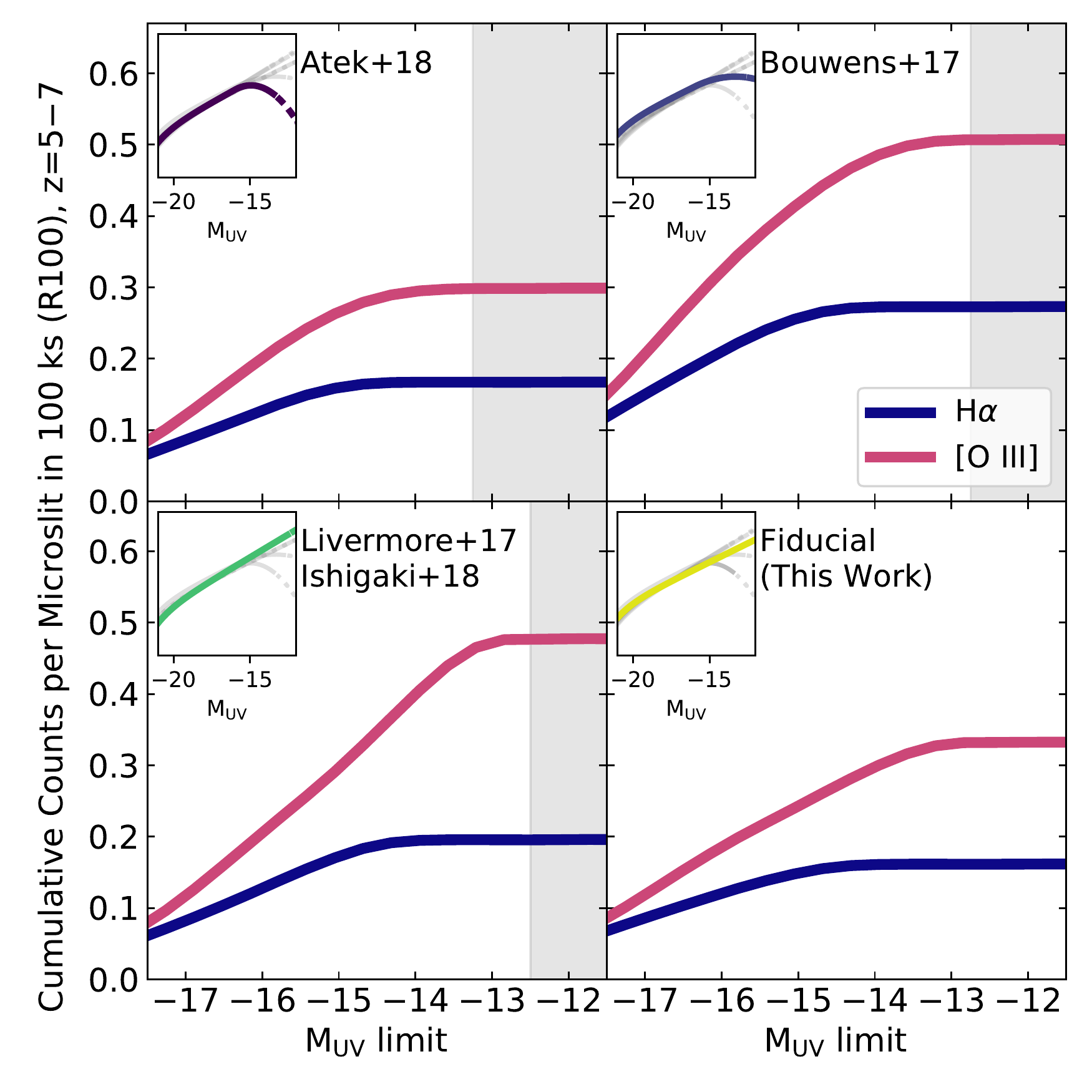}
\end{center}
\vspace{-0.5cm}
\caption{Cumulative number counts per microslit of predicted observable {\hbox{H $\alpha$}}- and \hbox{[O\,{\sc iii}]}-emitters at $z=5-7$ from a 100 ks R100 survey as a function of the faint \muv limit to which the luminosity function is probed.  Each panel shows the results for a different UV luminosity function: the top panels show the predictions for the \citet{2018MNRAS.479.5184A} and \citet{2017ApJ...843..129B} luminosity functions, which include a departure from a power-law at \muv $= -$16; the bottom panels show the predictions for \citet{2017ApJ...835..113L,2018ApJ...854...73I} and our fiducial model (Section \ref{sec:uvlf}) for power-law slopes to the faintest magnitudes.  The shaded grey regions denote the observational limit to each luminosity function.  Large samples of serendipitous \hbox{[O\,{\sc iii}]} and {\hbox{H $\alpha$}} emitters at these redshifts can, in conjunction with deep imaging data, constrain the shape and potential \muv cutoff of the UV luminosity function.}
\label{fig:z6counts}
\end{figure}

Results for {\hbox{H $\alpha$}} and \hbox{[O\,{\sc iii}]} counts as a function of the faint end of the UV luminosity function are shown in Figure \ref{fig:z6counts}.  Using the \citet{2017ApJ...843..129B}  luminosity function results in the largest predicted number of emitters (0.50 \hbox{[O\,{\sc iii}]} emitters and 0.27 {\hbox{H $\alpha$}} emitters per microslit) if the luminosity function extends to at least \muv $= -$12, with comparable results for the \citet{2017ApJ...835..113L} and \citet{2018ApJ...854...73I} luminosity function.  Using the \citet{2018MNRAS.479.5184A} luminosity function results in the fewest (0.30 \hbox{[O\,{\sc iii}]} emitters and 0.17 {\hbox{H $\alpha$}} emitters).

\begin{figure}
\begin{center}
\includegraphics[width=.4\textwidth]{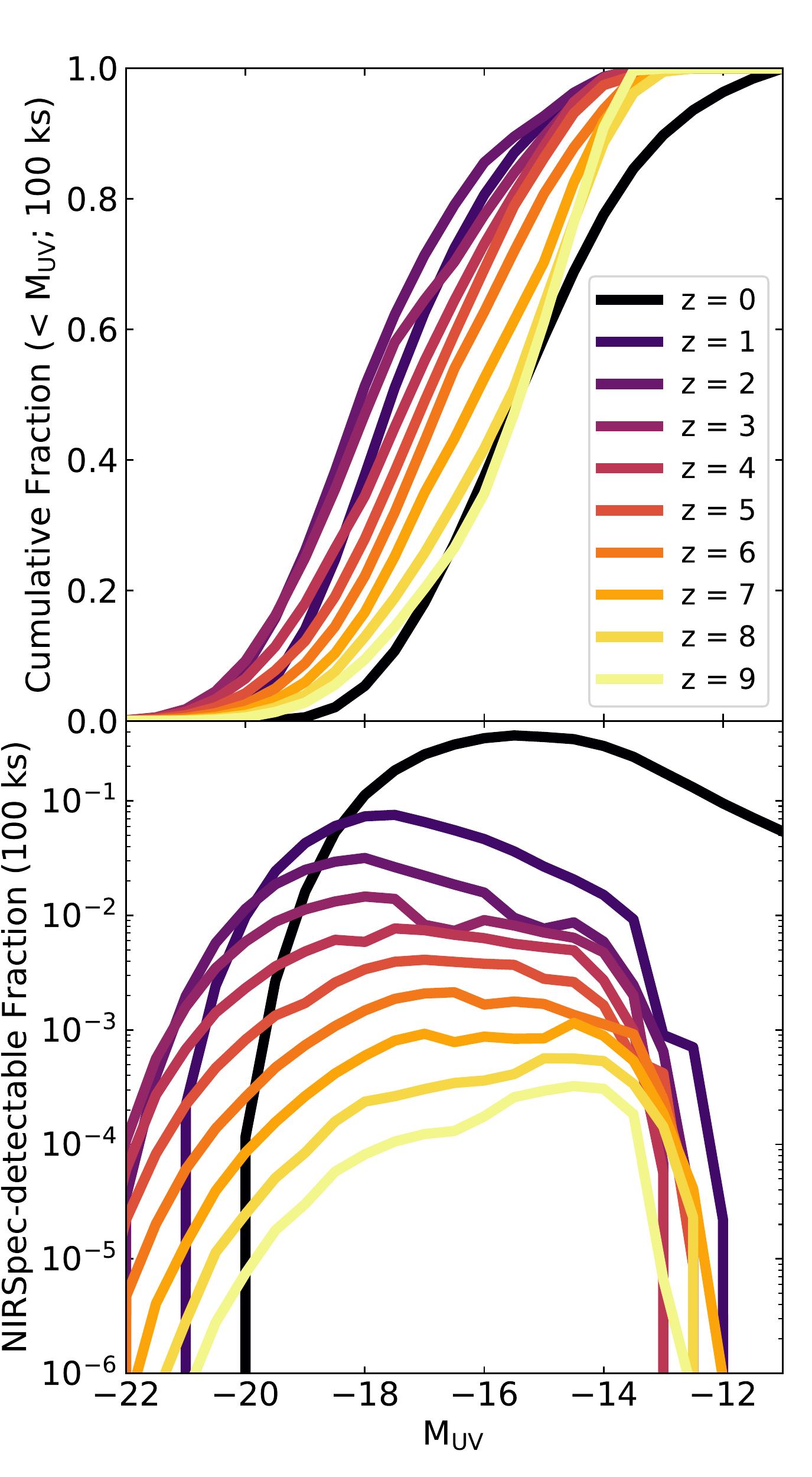}
\end{center}
\caption{(Top) Cumulative fraction of NIRSpec-detectable ($>$5-$\sigma$) galaxies in a 100 ks R100 observation as a function of \muv and $z$.  (Bottom) Total non-cumulative fraction of objects at a given \muv and $z$, based on our fiducial luminosity function, that have {\hbox{H $\alpha$}} or \hbox{[O\,{\sc iii}]} detectable in a 100 ks R100 observation.  While the lower-$z$ objects have a higher detectable fraction, the larger numbers of higher-$z$ galaxies mean that they dominate the total counts, as seen in Figure \ref{fig:tintU}.  At $z>5$, we also see that the majority of detections occur at \muv $\gtrsim$ $-$17.}
\label{fig:muvfrac}
\end{figure}

While UV luminosity functions obtained from deep NIRCam imaging will help to resolve this question \citep[e.g.][]{2018arXiv180309761Y}, NIRSpec will also be crucial in providing spectroscopic confirmations.  As the UV luminosity function needs to extend to at least \muv $\sim -$13 at $z\sim7-9$ in order for galaxies to reionize the universe \citep[e.g.][]{2013ApJ...768...71R}, serendipitous detections with NIRSpec could provide valuable constraints on the number counts of such galaxies considering not all \muv $> -$14 sources will be detectable even with the deepest NIRCam imaging (Figure \ref{fig:tintN}).  In fact, the \citet{2018ApJS..236...33W} JAGUAR catalogue does not contain any galaxies at $z>6$ with \muv $> -$14 that would be detectable in the `DEEP' NIRCam imaging (cf. their figure 25).  However, our model predicts NIRSpec confirmations for 0.07 objects per open microslit at $z>6$ with \muv $> -$14 in 100 ks at R100, and even 0.01 at $z>6$ with \muv $> -$13.  This is also visible in Figure \ref{fig:muvfrac}, where serendipitous counts with \muv $> -$14 become more prevalent at $z>4$ (top panel).

Deep (blind) NIRSpec exposures may therefore represent the most efficient way to find the faintest galaxies at high-$z$.  Confirming the UV magnitudes may be difficult when they are not detected in the broadband imaging, but a large enough sample could be stacked together to obtain an average measurement, as in \citet{2018ApJ...865L...1M}.  In fact, such confirmations might serve as an independent (spectroscopic-based) constraint on the faint end of the high-$z$ UV luminosity function.  At these faint magnitudes, NIRSpec detections are likely to make up only a low fraction of the total number of galaxies at high-$z$, $\sim$0.01 per cent (Figure \ref{fig:muvfrac}) highlighting the difficulty of studying the average properties of the sources likely responsible for reionization even with JWST.

\section{Conclusions}

In this work, we predict the number counts of emission line sources that are serendipitously detectable in the multi-object spectroscopy mode of JWST's NIRSpec instrument.  To do so, we develop a model that relates the fluxes and EWs of the {\hbox{H $\alpha$}} and \hbox{[O\,{\sc iii}]} emission lines to the \muv and redshift of the galaxy.  When coupled to a UV luminosity function that evolves smoothly with redshift, this predicts the line flux distributions of galaxies at a given redshift.  With no \textit{a priori} knowledge of the spatial location of these objects, we calculate the effective volume that NIRSpec probes as a function of the limiting detectable flux in a standard observing setup; even off-centre sources can have much of their flux enter through an open microslit.  This model therefore produces the following results:

\begin{itemize}
\item For integration times greater than 21 hours (75 ks) in the R100 mode, NIRSpec will serendipitously detect at least one {\hbox{H $\alpha$}} and/or \hbox{[O\,{\sc iii}]} emitter with $>$5-$\sigma$ significance per open microslit, i.e. there will be an equal or greater number of serendipitous sources than primary targets (Figure \ref{fig:tintU}).  At exposure times greater than 6 hours (20 ks), we predict at least 0.5 ($>$5-$\sigma$) emitters per microslit.  If we restrict to only {\hbox{H $\alpha$}} detections (e.g. if \hbox{[O\,{\sc iii}]} is much fainter than our model predicts), then we predict 0.67 detections per open microslit in 100 ks.
\item When considering objects that could be detected in broadband imaging, the expected number of serendipitous sources decreases at a fixed integration time depending on the type and depth of the imaging considered.  In this work, we consider HST-based imaging as in the CANDELS GOODS-S field as well as the combination of HST-based and JWST/NIRCam-based `DEEP' imaging in the UDF (Figures \ref{fig:tintC} and \ref{fig:tintN}).
\item The R100 mode of NIRSpec, with full spectral coverage from 0.6 to 5.3 $\mu$m, is the most efficient observing mode for detecting serendipitous sources at all exposure times (Figure \ref{fig:tint}).
\item The model predicts NIRSpec detections in some \muv $>-14$ galaxies, 92 per cent of which are \hbox{[O\,{\sc iii}]} emitters assuming our fiducial \hbox{[O\,{\sc iii}]} to {\hbox{H $\alpha$}} ratio, which might not be detectable even in the deepest NIRCam imaging to be taken with JWST in blank fields (Figures \ref{fig:tintN} and \ref{fig:muvfrac}).  Detections in such faint galaxies would represent an independent way to constrain the faint end of the high-$z$ UV luminosity functions (Figure \ref{fig:z6counts}).
\end{itemize}

Given the number of potential open microslits in the NIRSpec MSA, surveys intending to optimize number counts of emission line galaxies should strive to open as many as possible in any configuration.

\section*{Acknowledgements}
The authors would like to thank the anonymous referee for comments and suggestions that improved the quality of the manuscript.  This paper developed out of discussions with Jarle Brinchmann, Stefano Carniani, St\'ephane Charlot, Pierre Ferruit, Giovanna Giardino, Peter Jakobsen, Janine Pforr, and from the entire NIRSpec GA/MOS GTO team.  ECL acknowledges ERC Advanced Grant 695671 `QUENCH'.

\bibliographystyle{mnras} 
\bibliography{jwst}

\appendix
\section{UV Luminosity Functions}

\begin{table*}
\centering
\begin{tabular}{ccccl}
$<z>$ & $\alpha$ & log $\phi_{\star}$ & M$_{\star}$ & Reference \\
\hline
0.3 & $-$1.19$\pm$0.15 & $-$2.21$\pm$0.12 & $-$18.38$\pm$0.25  & \citet{2005ApJ...619L..43A}\\ 
0.5 & $-$1.55$\pm$0.21 &$-$2.77$\pm$0.23 & $-$19.49$\pm$0.37 & \citet{2005ApJ...619L..43A}\\
0.7 & $-$1.60$\pm$0.26 & $-$2.78$\pm$0.25 & $-$19.84$\pm$0.40  & \citet{2005ApJ...619L..43A}\\
1.0 & $-$1.63$\pm$0.45  & $-$2.94$\pm$0.29 & $-$20.11$\pm$0.45 & \citet{2005ApJ...619L..43A}\\
0.05 & $-$1.22$\pm$0.07 & $-$2.37$\pm$0.06 & $-$18.04$\pm$0.11 & \citet{2005ApJ...619L..15W}\\ 
1.7& $-$0.81$\pm$0.18 & $-$1.77$\pm$0.11 & $-$19.80$\pm$0.29 & \citet{2006ApJ...642..653S}\\ 
3.0& $-$1.43$\pm$0.13 & $-$2.77$\pm$0.11 & $-$20.90$\pm$0.18 & \citet{2006ApJ...642..653S}\\
4.0& $-$1.26$\pm$0.38 & $-$3.07$\pm$0.27 & $-$21.00$\pm$0.43 & \citet{2006ApJ...642..653S}\\
4.0 & $-$1.82$\pm$0.09 & $-$2.84$\pm$0.11 & $-$21.14$\pm$0.15 & \citet{2006ApJ...653..988Y}\\ 
4.7 & $-$2.31$\pm$0.64 & $-$3.24$\pm$0.58 & $-$21.09$\pm$0.64 & \citet{2006ApJ...653..988Y}\\
1.14 & $-$1.48$\pm$0.32 & $-$2.53$\pm$0.02 & $-$19.62$\pm$0.06 & \citet{2007ApJ...654..172D}\\ 
1.75 & $-$1.48$\pm$0.32 & $-$2.51$\pm$0.23 & $-$20.24$\pm$0.32 & \citet{2007ApJ...654..172D}\\
2.23 & $-$1.48$\pm$0.32 & $-$2.48$\pm$0.13 & $-$19.87$\pm$0.18 & \citet{2007ApJ...654..172D}\\
4.8 & $-$1.48$\pm$0.35 & $-$3.39$\pm$0.31 & $-$21.28$\pm$0.38 & \citet{2007MNRAS.376.1557I}\\ 
6.9 & $-$1.72$\pm$0.65 & $-$3.16$\pm$1.00 & $-$20.10$\pm$0.76 & \citet{2009ApJ...706.1136O}\\ 
5.0 & $-$1.66$\pm$0.06 & $-$3.03$\pm$0.09 & $-$20.73$\pm$0.11 &\citet{2009MNRAS.395.2196M}\\ 
6.0 & $-$1.71$\pm$0.11 & $-$2.74$\pm$0.12 & $-$20.04$\pm$0.12 &\citet{2009MNRAS.395.2196M}\\
2.3 & $-$1.73$\pm$0.07 & $-$2.56$\pm$0.09 & $-$20.70$\pm$0.11 & \citet{2009ApJ...692..778R}\\ 
3.05 & $-$1.73$\pm$0.13 & $-$2.77$\pm$0.13 & $-$20.97$\pm$0.14 & \citet{2009ApJ...692..778R}\\
1.7 & $-$1.27 & $-$2.66$\pm$0.15 & $-$19.43$\pm$0.36 & \citet{2010ApJ...720.1708H}\\ 
2.1 & $-$1.17$\pm$0.40 & $-$2.80$\pm$0.32 & $-$20.39$\pm$0.64 & \citet{2010ApJ...720.1708H}\\
2.7 & $-$1.52$\pm$0.29 & $-$2.81$\pm$0.32 & $-$20.94$\pm$0.53 & \citet{2010ApJ...720.1708H}\\
1.5 & $-$1.46$\pm$0.54 & $-$2.64$\pm$0.33 & $-$19.82$\pm$0.51 & \citet{2010ApJ...725L.150O}\\ 
1.9 & $-$1.60$\pm$0.51 & $-$2.66$\pm$0.36 & $-$20.16$\pm$0.52  & \citet{2010ApJ...725L.150O}\\
2.5 & $-$1.73$\pm$0.11 & $-$2.49$\pm$0.12 & $-$20.69$\pm$0.17 & \citet{2010ApJ...725L.150O}\\
3.1 & $-$1.65$\pm$0.12 & $-$2.75$\pm$0.11 & $-$20.94$\pm$0.14 & \citet{2010AA...523A..74V}\\ 
3.8 & $-$1.56$\pm$0.08 & $-$2.87$\pm$0.07 & $-$20.84$\pm$0.09 & \citet{2010AA...523A..74V}\\
4.8 & $-$1.65$\pm$0.09 & $-$3.08$\pm$0.08 & $-$20.94$\pm$0.11 & \citet{2010AA...523A..74V}\\
8.0 & $-$1.98$\pm$0.23 & $-$-3.37$\pm$0.28 & $-$20.26$\pm$0.32 & \citet{2012ApJ...760..108B}\\ 
0.13 & $-$1.05$\pm$0.04 & $-$2.15$\pm$0.03 & $-$18.12 &\citet{2012AA...539A..31C}\\ 
0.3 & $-$1.17$\pm$0.05 & $-$2.16$\pm$0.06 & $-$18.30$\pm$0.15 &\citet{2012AA...539A..31C}\\
0.5 & $-$1.07$\pm$0.07 & $-$2.18$\pm$0.06 & $-$18.40$\pm$0.10 &\citet{2012AA...539A..31C}\\
0.7 & $-$0.90$\pm$0.08 & $-$2.02$\pm$0.05 & $-$18.30$\pm$0.10 &\citet{2012AA...539A..31C}\\
0.9 & $-$0.85$\pm$0.10 & $-$2.05$\pm$0.05 & $-$18.70$\pm$0.10 &\citet{2012AA...539A..31C}\\
1.1 & $-$0.91$\pm$0.16 & $-$2.13$\pm$0.07 & $-$19.00$\pm$0.20 &\citet{2012AA...539A..31C}\\
1.5 & $-$1.09$\pm$0.23 & $-$2.39$\pm$0.09 & $-$19.60$\pm$0.20 &\citet{2012AA...539A..31C}\\
2.1 & $-$1.30 & $-$2.47$\pm$0.03 & $-$20.40$\pm$0.10 &\citet{2012AA...539A..31C}\\
3.0 & $-$1.50 & $-$3.07$\pm$0.03 & $-$21.40$\pm$0.10 &\citet{2012AA...539A..31C}\\
4.0 & $-$1.73 & $-$3.96$\pm$0.04 & $-$22.20$\pm$0.20 &\citet{2012AA...539A..31C}\\
2.2 & $-$1.47$\pm$0.23 & $-$2.74$\pm$0.32 & $-$21.00$\pm$0.55 & \citet{2012MNRAS.421.2187S}\\ 
7.0 & $-$1.90$\pm$0.15 & $-$2.96$\pm$0.21 & $-$19.90$\pm$0.26 & \citet{2013MNRAS.432.2696M}\\ 
8.0 & $-$2.02$\pm$0.23 & $-$3.35$\pm$0.38 & $-$20.12$\pm$0.43 & \citet{2013MNRAS.432.2696M}\\
7.0 & $-$1.87$\pm$0.18 & $-$3.19$\pm$0.26 & $-$20.14$\pm$0.42 & \citet{2013ApJ...768..196S}\\ 
8.0 & $-$1.94$\pm$0.23 & $-$3.50$\pm$0.34 & $-$20.44$\pm$0.41 & \citet{2013ApJ...768..196S}\\
2.0 & $-$1.74$\pm$0.08 & $-$2.54$\pm$0.15 & $-$20.01$\pm$0.24 & \citet{2014ApJ...780..143A}\\ 
8.0 & $-$2.08$\pm$0.30 & $-$-3.51$\pm$0.44 & $-$20.40$\pm$0.47 & \citet{2014ApJ...786...57S}\\ 
0.75 & $-$1.23$\pm$0.11 & $-$2.45$\pm$0.53 & $-$18.62$\pm$0.24 & \citet{2014ApJ...794L...3W}\\ 
1.25 & $-$1.19$\pm$0.11 & $-$2.83$\pm$0.38 & $-$18.97$\pm$0.22 & \citet{2014ApJ...794L...3W}\\
2.0 & $-$1.36$\pm$0.14 & $-$2.89$\pm$0.98 & $-$19.36$\pm$0.33 & \citet{2014ApJ...794L...3W}\\
3.0 & $-$1.36$\pm$0.13 & $-$3.00$\pm$0.63 & $-$20.45$\pm$0.23 & \citet{2014ApJ...794L...3W}\\
4.0 & $-$1.58$\pm$0.08 & $-$4.22$\pm$0.80 & $-$20.89$\pm$0.13 & \citet{2014ApJ...794L...3W}\\
5.0 & $-$1.57$\pm$0.11 & $-$4.22$\pm$1.81 & $-$20.83$\pm$0.17 & \citet{2014ApJ...794L...3W}\\
3.8 & $-$1.64$\pm$0.04 & $-$2.71$\pm$0.07 & $-$20.88$\pm$0.08 & \citet{2015ApJ...803...34B}\\ 
4.9 & $-$1.76$\pm$0.06 & $-$3.10$\pm$0.11 & $-$21.10$\pm$0.15 & \citet{2015ApJ...803...34B}\\
5.9 & $-$1.90$\pm$0.10 & $-$3.41$\pm$0.19 & $-$21.10$\pm$0.24 & \citet{2015ApJ...803...34B}\\
6.8 & $-$1.98$\pm$0.15 & $-$3.34$\pm$0.28 & $-$20.61$\pm$0.31 & \citet{2015ApJ...803...34B}\\
7.9 & $-$1.81$\pm$0.27 & $-$3.36$\pm$0.38 & $-$20.19$\pm$0.42 & \citet{2015ApJ...803...34B}\\
10.4 & $-$2.27 & $-$4.89$\pm$0.20 & $-$20.92 & \citet{2015ApJ...803...34B}\\
5.9 & $-$1.88$\pm$0.15 & $-$3.24$\pm$0.18 & $-$20.77$\pm$0.19 & \citet{2015MNRAS.452.1817B}\\ 
\hline
\end{tabular}
\caption{UV luminosity functions considered in this work.  In cases where asymmetric errors are reported, we include the average (symmetrized) errors; when no error is reported, we assume an error of 0.5.\label{tab:uvlf}}
\end{table*}

\begin{table*}
\contcaption{UV luminosity functions considered in this work.\label{tab:uvlfext}}
\centering
\begin{tabular}{ccccl}
$<z>$ & $\alpha$ & log $\phi_{\star}$ & M$_{\star}$ & Reference \\
\hline
4.0 & $-$1.56$\pm$0.06 & $-$2.85$\pm$0.06 & $-$20.73$\pm$0.09 & \citet{2015ApJ...810...71F}\\ 
5.0 & $-$1.67$\pm$0.06 & $-$3.05$\pm$0.08 & $-$20.81$\pm$0.13 & \citet{2015ApJ...810...71F}\\
6.0 & $-$2.02$\pm$0.10 & $-$3.73$\pm$0.20 & $-$21.13$\pm$0.28 & \citet{2015ApJ...810...71F}\\
7.0 & $-$2.03$\pm$0.21 & $-$3.80$\pm$0.34 & $-$21.03$\pm$0.44 & \citet{2015ApJ...810...71F}\\
8.0 & $-$2.36$\pm$0.47 & $-$4.14$\pm$0.96 & $-$20.89$\pm$0.91 & \citet{2015ApJ...810...71F}\\
9.0 & $-$2.02 & $-$3.60$\pm$0.23 & $-$20.1 & \citet{2015MNRAS.450.3032M}\\ 
1.3 & $-$1.56$\pm$0.04 & $-$2.63$\pm$0.09 & $-$19.74$\pm$0.18 & \citet{2016ApJ...832...56A}\\ 
1.9 & $-$1.72$\pm$0.04 & $-$2.82$\pm$0.11 & $-$20.41$\pm$0.20 & \citet{2016ApJ...832...56A}\\
2.6 & $-$1.94$\pm$0.06 & $-$3.26$\pm$0.11 & $-$20.71$\pm$0.11 & \citet{2016ApJ...832...56A}\\
7.0 & $-$1.91$\pm$0.27 & $-$3.43$\pm$0.13 & $-$20.33$\pm$0.42 & \citet{2016ApJ...820...98L}\\ 
8.0 & $-$1.95$\pm$0.42 & $-$3.52$\pm$0.75 & $-$20.32$\pm$0.38 & \citet{2016ApJ...820...98L}\\
9.0 & $-$2.17$\pm$0.42 & $-$4.15$\pm$0.19 & $-$20.45 & \citet{2016ApJ...820...98L}\\
1.7 & $-$1.33$\pm$0.03 & $-$2.17$\pm$0.05 & $-$19.61$\pm$0.07 & \citet{2016MNRAS.456.3194P}\\ 
1.9 & $-$1.32$\pm$0.03 & $-$2.15$\pm$0.04 & $-$19.68$\pm$0.05 & \citet{2016MNRAS.456.3194P}\\
2.25 & $-$1.26$\pm$0.04 & $-$2.12$\pm$0.05 & $-$19.71$\pm$0.07 & \citet{2016MNRAS.456.3194P}\\
2.8 & $-$1.31$\pm$0.04 & $-$2.27$\pm$0.05 & $-$20.20$\pm$0.07 & \citet{2016MNRAS.456.3194P}\\
3.8 & $-$1.43$\pm$0.04 & $-$2.69$\pm$0.07 & $-$20.71$\pm$0.10 & \citet{2016MNRAS.456.3194P}\\
5.9 & $-$1.91$\pm$0.03 & $-$3.19$\pm$0.04 & $-$20.94 & \citet{2017ApJ...843..129B}\\ 
6.0 & $-$2.10$\pm$0.03 & $-$3.65$\pm$0.04 & $-$20.83$\pm$0.05 & \citet{2017ApJ...835..113L}\\ 
7.0 & $-$2.06$\pm$0.04 & $-$3.67$\pm$0.04 & $-$20.80$\pm$0.05 & \citet{2017ApJ...835..113L}\\
8.0 & $-$2.01$\pm$0.08 & $-$3.77$\pm$0.13 & $-$20.72$\pm$0.16 & \citet{2017ApJ...835..113L}\\
7.0 & $-$1.98$\pm$0.10 & $-$3.43$\pm$0.21 & $-$20.74$\pm$0.21 & \citet{2018MNRAS.479.5184A}\\ 
6.5 & $-$2.15$\pm$0.07 & $-$3.78$\pm$0.15 & $-$20.89$\pm$0.15 & \citet{2018ApJ...854...73I}\\ 
8.0 & $-$1.96$\pm$0.17 & $-$3.60$\pm$0.23 & $-$20.35$\pm$0.25 & \citet{2018ApJ...854...73I}\\
9.0 & $-$1.96 & $-$3.88$\pm$0.10 & $-$20.35 & \citet{2018ApJ...854...73I}\\
10.0 & $-$1.96 & $-$4.60$\pm$0.18 & $-$20.35 & \citet{2018ApJ...854...73I}\\
10.0 & $-$2.28$\pm$0.32 & $-$4.89$\pm$0.27 & $-$20.60 & \citet{2018ApJ...855..105O}\\
\hline
\end{tabular}

\end{table*}

\bsp	
\label{lastpage}
\end{document}